\documentclass[12pt]{iopart}
\usepackage{iopams}  
\usepackage{graphicx}
\usepackage{bm}

\newcommand{\aeq}{\begin{equation}}
\newcommand{\eeq}{\end{equation}}
\newcommand{\aeqn}{\begin{eqnarray}}
\newcommand{\eeqn}{\end{eqnarray}}
\newcommand{\aeqs}{\begin{equation*}}
\newcommand{\eeqs}{\end{equation*}}
\newcommand{\aeqns}{\begin{eqnarray*}}
\newcommand{\eeqns}{\end{eqnarray*}}
\DeclareOldFontCommand{\bf}{\normalfont\bfseries}{\mathbf}
\DeclareOldFontCommand{\rm}{\normalfont\rmfamily}{\textrm}
\newcommand{\vecb}[1]{{\textbf #1}}

\newcommand{\vpar}{{v_\parallel}}

\newcommand{\fpar}[2]{\frac{\partial{#1}}{\partial{#2}}}

\newcommand{\fder}[2]{\frac{d{#1}}{d{#2}}}

\def \d {\mathrm{d}}

\newcommand{\Alfv}{Alfv\'en}

\newcommand{\be}{\begin{equation}}
\newcommand{\ee}{\end{equation}}
\newcommand{\ba}{\begin{eqnarray}}
\newcommand{\ea}{\end{eqnarray}}
\newcommand{\bas}{\begin{eqnarray*}}
\newcommand{\eas}{\end{eqnarray*}}
\newcommand{\vphi}{\varphi}

\newcommand{\df}{
{\mathrm d}
}

\begin{document}

\title[Particle-in-cell methods in edge plasma physics: the PICLS code]{Particle-in-cell methods in edge plasma physics: the PICLS code}

\author{\underline{A.~Bottino}$^{1}$, A.~Stier$^{1}$, M.~Boesl$^{2}$,
  T.~Hayward-Schneider$^{1}$, A.~Bergmann$^{1}$, D.~Coster$^{1}$,
  S.~Brunner$^{3}$, G.~Di~Giannatale$^{3}$ and L.~Villard$^{3}$}

\address{$^1$Max-Planck-Institut f\"ur Plasmaphysik, 85748 Garching, Germany \\
$^2$Spark e-Fuels GmbH, Berlin, Germany\\
$^3$Swiss Plasma Center, Ecol e Polytechnique F\'ed\'erale de Lausanne, Switzerland\\
}
\ead{alberto.bottino@ipp.mpg.de}
\vspace{10pt}
\begin{indented}
\item[]August 2024
\end{indented}

\begin{abstract}
  Over the past decades, multiple gyrokinetic codes have shown to be able to
simulate turbulence and associated transport in the core of Tokamak devices. However, their
application to the edge and scrape-off layer (SOL) region presents significant challenges.
To date, only few codes and models have been adapted to SOL/edge conditions.
To further study the SOL region in particular, with its steep temperature and density gradients
as well as large fluctuation amplitudes, the full-f particle-in-cell code PICLS has been
developed. PICLS is based on a full-f gyrokinetic model with linearised field equations,
considers kinetic electrons and uses logical sheath boundary conditions.  In the past, PICLS was verified
by applying it to a well-studied 1D parallel transport problem during an edge-localized
mode in the SOL under both collisionless and collisional
conditions, for which a Lenard-Bernstein collision operator was implemented.
PICLS recently was extended towards three spatial dimensions to study turbulence in
open-field-line regions in slab and closed-field-line toroidal geometries.
In this work, we will focus on the models and methods we used for extending the code
towards three spatial dimensions, including validation efforts and comparisons with
other existing codes in closed-field-line geometry.
\end{abstract}

%
%
%
%
%

\section{Introduction}
Plasma theory and simulation of Tokamak plasmas advanced significantly during the past decades.
In particular, gyrokinetic simulations have been proven to largely reproduce turbulence transport
in the core, the innermost and hottest region of the plasma, where fusion power will be produced.
The magnetic flux surfaces in the core region are nested and closed. The edge region is the subsequent region
further towards the plasma wall, which also is characterized by closed flux surfaces.
It is a rather thin layer and compared to the core often exhibits steep pressure gradients \cite{Zweben07}.
Once crossing the last closed flux surface (LCFS or Separatrix) an often even thinner layer of open field lines begins,
the scrape-off layer (SOL). Here, the open field lines wind around toroidally several times before they hit the divertor target.
Thus, in the SOL particles are lost to the machine surface.
In general, the greatest part of the plasma lies within the core and edge region. This is required, to achieve high enough confinement for the fusion reaction.
Particles that would be confined by the closed flux surfaces, however, can be transported across the surfaces by turbulence and collisions. 
In the SOL, plasma that is transported out of the confined region, in general is lead to the divertor targets before it can reach the wall. This region is highly important for regulating the impurity level and confining the plasma. Plasma dynamics in the SOL are defined by a balance of plasma outflow from the core, turbulent transport across magnetic field lines, parallel flow and losses towards the divertor targets \cite{stangeby2000plasma,Ricci15,Mosetto13}. Typical characteristics for the SOL are low plasma density and temperature ($T_e\sim10-100$eV, due to the plasma-wall interaction \cite{Zweben07}), high collisionality and large-scale turbulence of amplitudes of order of unity compared to the background. Experimentally measured profiles confirmed these large amplitude fluctuations and the exponential decay of radial profiles.
The complex physics in the SOL region also increases the complexity for plasma
modeling and simulation \cite{Cohen94}. For example, in the parallel direction along the
field lines, electrons hit the divertor faster than ions, due to their higher mobility. The plasma establishes a thin layer of net
positive charge of a few Debye lengths (due to Debye shielding), directly at
the plasma-wall interface, the so-called Debye sheath \cite{stangeby2000plasma}. This layer builds up a potential barrier that
accelerates incident ions into the wall and repels incident electrons (only
the fastest electrons can overcome the potential barrier). The particle fluxes
of both species are kept approximately equal, thus the plasma is maintained
quasi-neutral. The SOL plasma properties and the particle and energy flux
towards the surface are significantly influenced by this sheath. In general plasma/wall 
boundary conditions are very challenging, due to the very fast dynamics of the plasma at the interface, combined with complicated geometries, requiring body-fitted unstructured meshes or remeshing strategies which can be time consuming.
On the other hand, penalization techniques \cite{ISOARDI20102220}  have been proven to be an efficient tool to model
efficiently the plasma/limiter interaction.
Moreover, neutral gas dynamics also plays a major role in the SOL region, by decreasing the 
temperature and increasing the collisionality.  

Apart from that, the size of plasma structures ranges from the gyroradius to machine size with strong gradients perpendicular and along the field lines. Also time scales vary in a wide range between the gyromotion and turbulence time scales. Fluctuations are of the order of unity compared to the background, which means that in simulations a {\em linearization} of quantities (i.e., splitting the distribution function in a constant background and a small perturbation $f=f_0 + \delta f $) as often applied in the plasma core is not valid anymore. And finally, also the divertor geometry adds up to the increase in complexity.\\
Although there exist fully six-dimensional kinetic code developments \cite{Tskhakaya09, Manfredi11}, large time and length scale Tokamak simulations that solve the Vlasov–Maxwell or Vlasov–Poisson equations require processing and memory capabilities that are far beyond what nowadays supercomputers can provide. Even now, when we approach exascale computing, the simulations are still way too costly. To cope with these computational restrictions, physical approximations are required to reduce the numerical calculation efforts.\\  
Simulations within the SOL have been and are still primarily performed with the help of fluid- \& gyrofluid-based codes, e.g.~\cite{Schneider92, Schneider06, Rozhansky09, Stegmeir18, Zholobenko19, Scott03}. The key advantage of fluid codes is their computationally relatively low effort, compared to kinetic approaches. Nevertheless, kinetic effects play an important role in edge plasmas with steep gradients and including the open-field region \cite{Takizuka_2017}. Therefore, there are major efforts to apply gyrokinetic first-principle codes to SOL and edge plasma. The underlying physical model is a 5D description of low-frequency plasma dynamics reduced from the initial 6D kinetic equations \cite{Tronko16, Sugama00, Brizard20004816}.
In the plasma core gyrokinetic simulations are nowadays widely used and produce comparable turbulence results across various codes.
But due to the extremely rich physics in the plasma edge/SOL, only few codes have been adapted to simulate the plasma periphery.
To only name a few challenges that are especially relevant for gyrokinetic models in this region: large amplitude fluctuations, sheath boundaries, wide range of space and time scales, atomic physics, etc. As a result, code developments in the SOL need to specifically address these complications.
Currently, one of the most advanced gyrokinetic code is probably XGC, which gained popularity with its calculation of the ITER scrape-off layer width \cite{Chang17} and it was recently applied to model tungsten transport in the pedestal and turbulence transport in the presence of resonant magnetic perturbations (see Refs.~\cite{DominskiPoP2024_1,DominskiPoP2024_2, Hager_2019} and references therein).\\
Most commonly, there are two basic methods applied to numerically solve the
gyrokinetic system, particle-in-cell (PIC) methods (e.g.~\cite{Bott15, Lee83,
  KU2016467}) and continuum methods (e.g.~\cite{Jenko01, Hakim2020}).
In the continuum case (also called ``Eulerian''), the 5D gyrokinetic equations
are discretized on a fixed phase-space mesh and solved on this mesh. The
resulting partial differential equations are then solved with numerical
methods like finite-volumes, finite-elements, etc. Equal to the PIC approach,
in continuum codes the 3D fields are calculated by using grid-based
algorithms. In a PIC code Monte Carlo methods are applied that approximate integrals over phase space (moments of the particle distribution function) by a finite number of markers \cite{Krommes12}. In a more tangible picture, the markers can be described as ``superparticles'' that comprise many physical particles. The number of markers used in practice is much lower than the physical particle number \cite{Tskhakaya07}.
In a very simplified picture, the PIC algorithm applied on gyrokinetic plasma simulations advances these particles according to the Euler Lagrange equations, with the help of a force field which is set up by a charge/current distribution (\cite{Hockney88},\cite{Birdsall04}). 
More specifically, markers are initially distributed in the phase space and according to their positions, the 3D field equations are solved from this charge distribution on a fixed grid. Then the resulting fields are
interpolated back to the marker positions and using these fields the markers are advanced to their new position according to the gyrokinetic Euler-Lagrange equations. This procedure is then repeated again and again for the new distributions. One of the main challenges of PIC algorithms is their inherent statistical noise in the distribution function, due to their Monte Carlo sampling approach. For $N$ markers, this error scales with $1/\sqrt{N}$. This effect is extremely relevant for the validity of results and thus several techniques for noise reduction have been developed, e.g.~\cite{Garbet10, Krommes07, Bottino07}. Another often discussed problem resulting from statistical noise is the so-called Amp\`ere's law cancellation problem in electromagnetic simulations, where two large terms in the Amp\`ere's law cancel out numerically at average to high plasma $\beta$ and small perpendicular wave numbers. But since its discovery, various mitigation techniques were introduced with convincing results \cite{Hatzky07, Mishchenko04, Kleiber16, Mishchenko17} and nowadays it is not seen as a showstopper anymore.\\
Numerical models and challenges behind PIC and continuum codes for edge/SOL applications are
quite different (see e.g.~\cite{Idomura14, Xu07, Scott10GEM, Lee18cogent, Hakim14,Shi17,Shi18,Michels20}). Thus, to be able to perform reliable quantitative studies, it is important to develop both further and use cross-checks between both methods
for validation. In this paper we present the gyrokinetic full-f particle-in-cell code for open field lines in the SOL, called PICLS.
PICLS is a gyrokinetic Particle-in-Cell code for simulations in the scrape-off layer, or open field line regions.
The code was initially designed to study sheath boundary models and heat and particle fluxes from the plasma towards the wall,
but gets constantly extended towards more complicated open field line geometries.
The PICLS code was already introduced in \cite{Boesl2019, Boesl2020}, but
limited to the a single spatial dimension (1D).
While a 1D code can be used to model the parallel transport in open field lines, the study of the cross-field transport in the SOL requires a 3D model.
For magnetic fusion devices, such as tokamaks, in the SOL the propagation of
blobs (also called plasma filaments) can convectively transport heat, particles,
momentum and current across magnetic field lines and thus lead to a highly
discontinuous cross-field transport (see e.g.~\cite{Zweben07}). 
It is believed that in tokamak devices across the blob cross-section the curvature
and $\nabla \vecb{B}$ forces establish a charge-separated dipole
potential. This mainly causes an outward radial propagation of the blob, due
to convective $\vecb{E}\times\vecb{B}$ transport. For advanced fusion devices,
such as ITER and beyond, the balance between parallel and cross-field
transport in the SOL is decisive for how heat and particles are exhausted
(\cite{Loarte07}). Thus, for future performance predictions of fusion devices,
modeling of these turbulent transport phenomena in the SOL region is
important. Therefore, we had to extend our previous 1D code to a 3D version
which can cope with different geometries and includes magnetic curvature and
$\nabla \vecb{B}$ terms. In this work, the 3D version of the PICLS code is
presented, together with some validation efforts and comparisons with existing codes.\\
Just as a short side note, this is also the reason for its name ``PICLS'', a combination of ``PIC'' for Particle-In-Cell and ``LS'' for Logical Sheath \cite{PARKER1993}. So far, a rather simple 1D (one spatial domain) case and a more advanced 3D case are implemented in PICLS. 

\section{The PICLS code}
\subsection{The physical model}
In general, the physical model behind PICLS is derived from a gyrokinetic Lagrangian (see e.g. \cite{Sugama00}), using the Hamiltonian for the particle species $p$
\aeq\label{electrostatic_hamiltonian}
H_p = m_p\frac{\vpar^2}{2}+\mu B+ e_p J_{p,0}\phi-\frac{m_pc^2}{2B^2}|\nabla_\perp \phi|^2,
\eeq
with the velocity variables $\vpar$ (parallel velocity), $\mu=m_p v_\perp^2/(2B)$ (magnetic moment), the magnetic field strength $B$, the mass $m_p$ and charge $e_p$; $c$ is the speed of light. The gyroaveraging operator is indicated as $J_{p,0}$. In the following, variables with a perpendicular subscript lie within the plane perpendicular to the magnetic background field $\vecb{B}$, whereas variables with a parallel subscript lie along the magnetic background field.
Details about the specific derivation of the PICLS model can be found in \cite{Tronko18}.\\
The corresponding Euler-Lagrange equations are 
\aeqn
\dot{\vecb{R}} & =
\vpar\frac{\vecb{B^*}}{B^*_\parallel}+\frac{c}{e_pBB^*_\parallel}\vecb{B}\times
\nonumber \\
& \left[\mu
  \nabla B + e_p \nabla J_{p,0} \phi-\frac{m_pc^2}{2B^2}\nabla |\nabla_\perp \phi|^2 +\frac{m_pc^2}{B^3}\nabla B |\nabla_\perp \phi|^2\right],\nonumber\\
\dot{\vpar } & = -\frac{\vecb{B^*}}{B^*_\parallel}\frac{1}{m_p}\cdot\left[\mu \nabla B + e_p \nabla J_{p,0}\phi -\frac{m_pc^2}{2B^2}\nabla |\nabla_\perp \phi|^2 +\frac{m_pc^2}{B^3}\nabla B |\nabla_\perp \phi|^2\right],\nonumber\\
\dot{\mu }& = 0.
\eeqn
The Vlasov equation is
\aeq
\frac{d}{dt} f = \fpar{f}{t}+\dot{\vecb{R}}\cdot\nabla f + \dot{\vpar}\fpar{f}{\vpar} = C,
\eeq
where $C$ indicates a generic collision operator and upper dot correspond to Lagrangian derivatives.
The polarization (Poisson) equation, in the long wavelength limit,  has the form of a nonlinear elliptic equation
\aeq
-\sum_p \nabla_\perp \frac{n_\textrm{i} m_pc^2}{B^2} \nabla_\perp \phi=\sum_p \int \d W  e_pJ_{p,0} f.
\eeq
However, in this work we considered a simpler model, for which 
the so-called linearized polarization approximation has been used (see \cite{Bott15} end references therein).
Nevertheless, ion finite-Larmor radius effects (FLR) are retained in both the polarization density and in the gyroaverage operator.
The full derivation of the model equations can be found in \cite{Boesl2019}.
In this simpler case, the Euler-Lagrange equations do not contain second order terms 
\aeqn
\dot{\vecb{R}} & = \vpar\frac{\vecb{B^*}}{B^*_\parallel}+\frac{c}{e_pBB^*_\parallel}\vecb{B}\times\left[\mu
  \nabla B + e_p \nabla J_{p,0} \phi\right],\nonumber\\
\dot{\vpar } & = -\frac{\vecb{B^*}}{B^*_\parallel}\frac{1}{m_p}\cdot\left[\mu \nabla B + e_p \nabla J_{p,0}\phi \right],\nonumber\\
\dot{\mu } & = 0.
\eeqn
The polarization equation is now a linear elliptic equation because the gyrocenter density $n_{\textrm{i}}$ is replaced by
the equilibrium background density $n_{\textrm{i},0}$
\aeq
-\sum_p \nabla_\perp \frac{n_{\textrm{i},0} m_pc^2}{B^2} \nabla_\perp \phi=\sum_p \int \d W  e_pJ_{p,0} f.
\eeq
In addition to this, collisions, while being included in PICLS, are not considered in this work. The Vlasov equation becomes
\aeq
\frac{d}{dt} f = \fpar{f}{t}+\dot{\vecb{R}}\cdot\nabla f + \dot{\vpar}\fpar{f}{\vpar} = 0.
\eeq
The total conserved energy in the electrostatic model is \cite{Dubin83,Tronko18}
\aeqn
\mathcal{E}_\textrm{tot} &=& \sum_p \int \d W \d V H_p f_p \\
&=& \mathcal{E}_\textrm{k}+\mathcal{E}_\textrm{f}=\sum_p \int \d W \d V H_{p,0} f_p + \sum_p \frac{1}{2} \int \d W  \d V e_p J_{p,0} \phi f_p,
\eeqn
with $H_{p,0}=\frac{1}{2}m_p\vpar^2+\mu B$ in the kinetic part.\\

\subsection{Normalization}
In PICLS we implemented two different sets of normalization schemes. In the first set, mostly used for debugging,
the code uses the standard CGS units of the physical quantities. In the second one, all
the equations are normalized using input reference values before numerical solutions are calculated.
Given a dimensional quantity $x$, we split it into a dimensional constant
$\bar{x}$ and a dimensionless quantity $\tilde{x}$, as illustrated if Table~\ref{tab1}.

\begin{table}[htb]
\caption{\label{tab1}PICLS geometric setup and initial perturbation for the four simulations of Fig.~\ref{fig:genexcomp}. Each of the four
  simulations is characterized by different values of $q$. The poloidal and toroidal mode numbers are selected such that $k_\parallel=1$
  in all cases. Consequently, in each simulations the value of $m$ has been adapted to the corresponding $q$:  $m=5$ for $q=5/3$,
  $m=10$ for $q=10/3$, $m=15$ for $q=15/3$ and $m=20$ for $q=20/3$.}
\begin{indented}
\item[]\begin{tabular}{@{}lll}
\br
Quantity &Label 2&Normalisation\\
\mr
  time & $t$ & $\tilde{t}\bar{t} $ \\ 
    length & $l$ & $\tilde{l}\bar{l} $ \\
    mass & $m$ & $\tilde{m}\bar{m} $ \\
    charge & $q$ & $\tilde{q}\bar{q} $ \\
    Magnetic field & $B$ & $\tilde{B}\bar{B} $ \\
    Temperature & $T$ & $\tilde{T}\bar{T} $ \\
    velocity & $v$ & $\tilde{v}\bar{v} $ \\
    magnetic moment & $\mu$ & $\tilde{\mu}\bar{\mu} $ \\
    electrostatic pot. & $\phi$ & $\tilde{\phi}\bar{\phi} $ \\
    electromagnetic pot. & $A_{\parallel}$ & $\tilde{A_{\parallel}}\bar{A_{\parallel}}$ \\
    current density & $j_\parallel$ & $\tilde{j}_\parallel\bar{j}_\parallel $\\
    charge density & $\rho$ & $\tilde{\rho}\bar{\rho}$ \\
\end{tabular}
\end{indented}
\end{table}

Only five of the dimensional quantities of Table~\ref{tab1} are independent. In PICLS, the 
independent normalization quantities are chosen to be, and their normalization
is described in Table~\ref{tab2}.

\begin{table}
  \caption{\label{tab2}Normalization quantities in PICLS.}
   \begin{indented}
   \item[]\begin{tabular}{@{}lll}
     \br
     Parameter & Normalisation & Description \\
     \mr
    $\bar{B}$ & $B_0$ & magnetic field strength\\
    $\bar{m}$ & input or $m_e$ & mass\\
    $\bar{q}$ & input or $|q_e|$ & charge \\ 
    $\bar{T}$ & $T_e$ at $r_{\textrm ref}$ & temperature\\
    $\bar{n}$ & $N_{phys}/V$ & density\\
  \end{tabular}
\end{indented}
\end{table}

Where $N_{phys}$ is the total number of physical electrons, $V$ is the plasma volume and $r_{\textrm ref}$ an arbitrary chosen
radial reference point. The PICLS output is always in CGS units.\\

\subsection{Discretization}
\begin{figure}[htb]
\centering
\includegraphics[width=\textwidth]{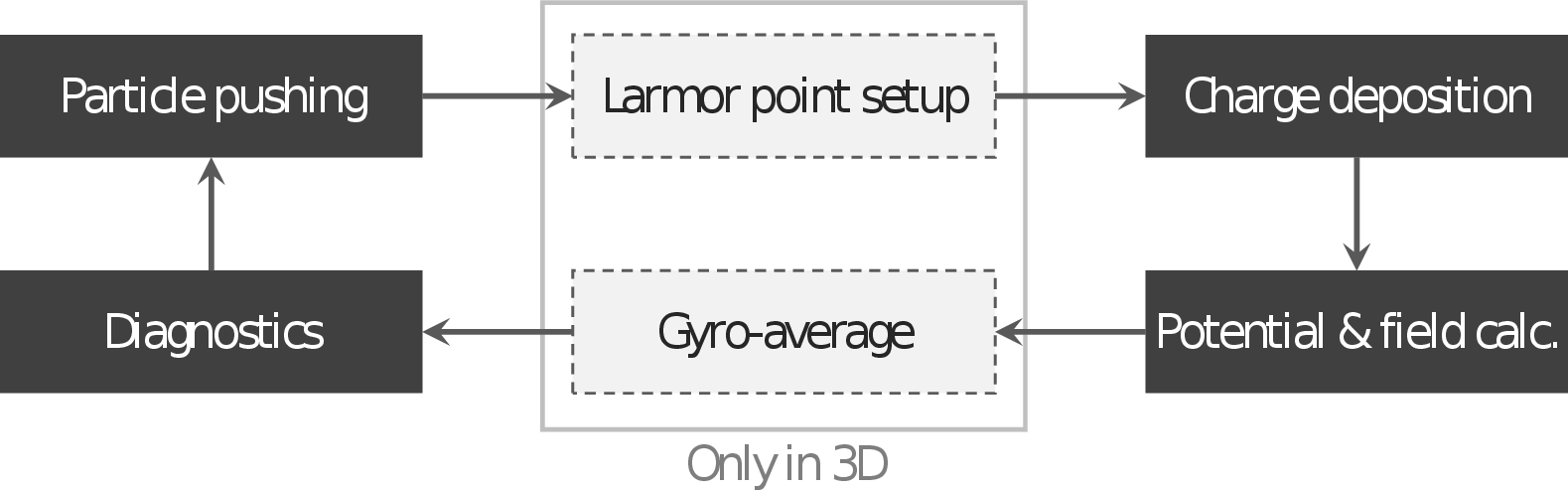}
\caption{Schematic representation of main procedures applied at each time step of simulation. Procedures with bright gray background are only applied in 3D case if Larmor radius effects are considered. Other procedures are used for each dimensionality (1D and 3D).}
\label{fig:timestepalgo}
\end{figure}

The most important difference between the 1D and 3D version of PICLS is the introduction of the gyroaverage operator.
As in the ORB5 code, the gyroaverage is performed in real space with a discrete average on the gyroring of each single marker.
The standard PIC cycle, based on particle pushing and charge deposition is therefore modified
by setting up ``Larmor points'' on the Larmor radius around the gyrocenter before the charge deposition and
by averaging across these Larmor points directly after the field calculation, following Ref.~\cite{lantiCPC2020}.
The PICLS computational cycle is summarized in Fig.~\ref{fig:timestepalgo}.
The number of Larmor points used to calculate the average depends on the thermal velocity of the particles and thus on the size of the gyroradius \cite{HatzkyJPP2019, DominskiPoP2018}.
A larger gyroradius requires a higher number of Larmor points, since the
points are distributed further and thus the probability for the fields to
significantly vary their value along the gyroring, significantly increases.\\
These procedures are repeated again and again in so-called timesteps via a
Runge-Kutta 4th order time integrator and for each time step system
diagnostics can be extracted.
Note that PICLS also allows for a drift-kinetic model for ions and
electrons. In this case, the only FLR effect retained is the polarization
density, while all the gyroveraged quantities are replaced by their value at
the gyro-center position, details can be found in \cite{Bott15}. In all the
simulations presented in this paper, a drift-kinetic model is used for the
electrons and the electron polarization density is neglected in the Poisson equation.

\subsection{Initialization of the markers}
\begin{figure}
\centering
\includegraphics[width=\textwidth]{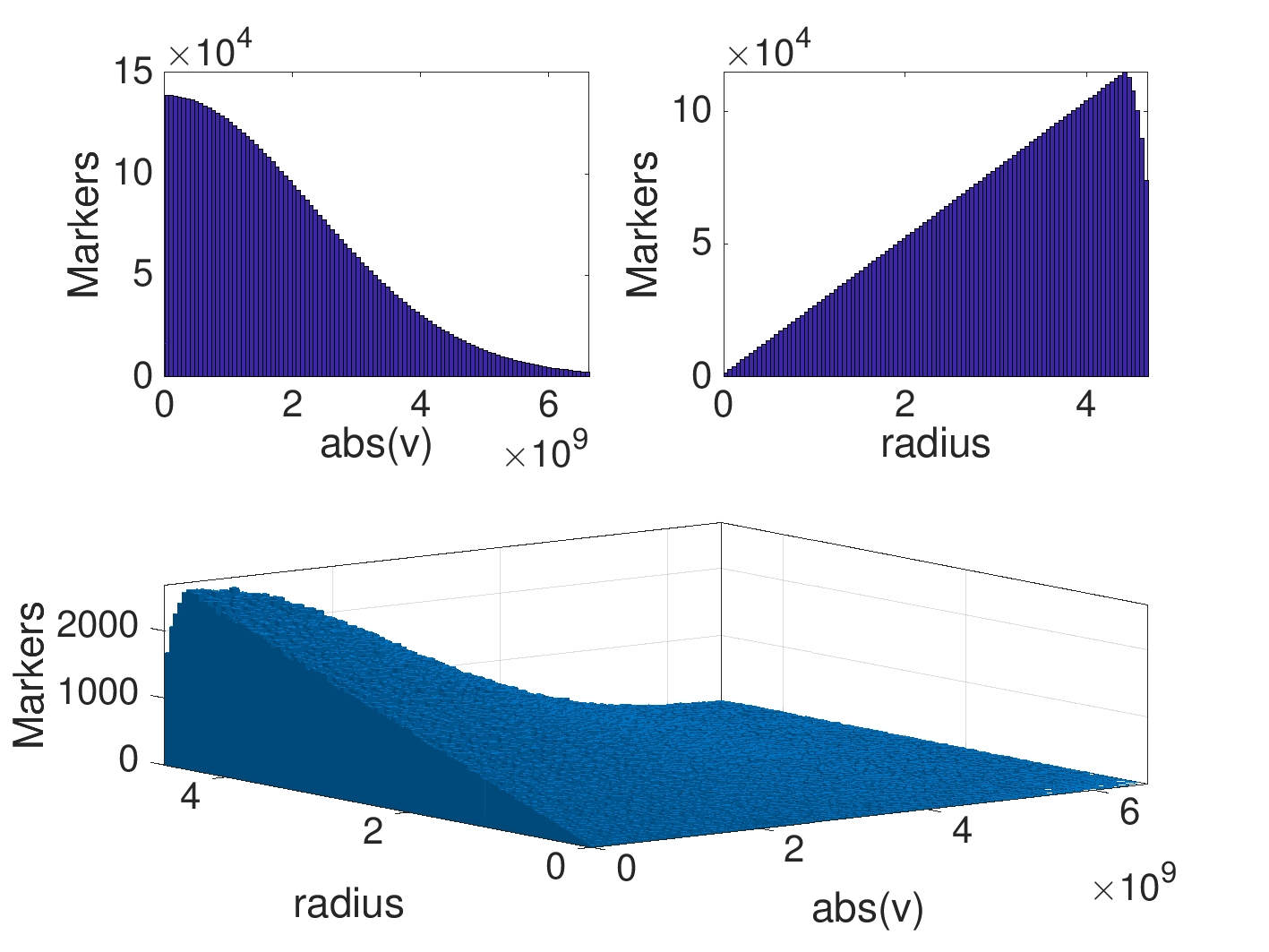}
\caption{Example of Maxwellian loading.
  Histograms in velocity and radius for a subset of 6 million test particles, loaded assuming
  a Maxwellian probability distribution function with constant $T$ and $n$
  profiles. The analytically prescribed Maxwellian function is correctly reproduced by the markers.}
\label{fig:maxwellian_loading}
\end{figure}

An important difference between PICLS and other existing PIC codes like ORB5 is that here the entire gyrocenter distribution function $f(\vecb{R})$
is represented by discrete markers. In the full-f representation, where the whole distribution will be simulated, the particle
distribution function thus can be expressed as
\aeq\label{discretef}
f(\vecb{R},\vpar,\mu,t)=\sum_{n=1}^{N}w_n(t) \delta(\vecb{R}-\vecb{R}_n(t))\delta(\vpar-\vpar_n(t))\delta(\mu-\mu_n),
\eeq
with $N$ the number of markers, $w_n$ the marker weights, $\vecb{R}_n$ their position, $\vpar_n$ their parallel velocity and $\mu_n$ their (constant) magnetic moment. With the definition for the initial number of physical particles, $N_\textrm{phys}=\int n_0(\vecb{R})\d \vecb{R}$, the weights in our case are uniformly initialized with 
\aeq
w_n=\frac{1}{N}, 
\eeq
for all markers. The weights for the full-f case are constant and thus do not change over time:
\aeq
\fder{}{t}w_n=0. 
\eeq
More details about the exact definition of the marker weights for PICLS and ORB5 can be found in Ref.~\cite{Bottino2022}.
In the 3D version of PICLS, markers are loaded in phase-space using importance sampling. In all the simulations presented in this work, a
Maxwellian loading has been used, proportional to the physical equilibrium distribution function, chosen to be a Maxwellian described by
density and temperature profiles, function of the radius only. As a result, the marker weights are all the same and normalized in a way
that their sum corresponds to one. An example of Maxwellian loading is illustrated in the histograms of Fig.~\ref{fig:maxwellian_loading}, where
the particle distribution in velocity has been reconstructed after the initial loading, by binning a set of test particles (6M) in radius
and $|v|$.

\subsection{3D finite element solver}
PICLS uses a Finite Element (FE) discretization scheme with B-splines as basis functions.
Solving the 3D electric potential with finite elements in all three dimensions can be computationally extremely costly,
depending on the number of grid cells in each direction ($n_x, n_y, n_z$) or ($n_r,n_\theta,n_\vphi$). However, if the regarded problem is periodic in one of the three dimensions,
a Discrete Fourier Transform (DFT) can be applied to the B-spline coefficients in the  periodic dimension. Depending on the size of the problem,
this can reduce the computational cost of the simulation significantly. The remaining two dimensions can still be calculated based on a 2D finite element field solver.  This Fourier enhanced finite element solver of PICLS has been described in detail in a dedicated paper \cite{Stier2024}.
Assuming $\vphi$ being the periodic direction, the result is that the large matrix corresponding to the 3D Poisson equation is replaced by a set of $n_\vphi$ smaller matrices equations (2D), making the solver straightforwardly parallelizable,
\aeq\label{Msolver_eq}
\sum_{j'k'} \phi_{j'k'}^{(n)}\,A_{j'k'jk}=\frac{b_{jk}^{(n)}}{M^{(n)}},
\eeq
with
\aeq
A_{j'k'jk}=\int\,N(s,\theta)  
\nabla_\perp(\Lambda_{j'}(s) \Lambda_{k'}(\theta)) 
    \nabla_\perp(\Lambda_j(s)
\Lambda_k(\theta)) \, J(s,\theta) \, \df s \, \df \theta
\eeq
and
\aeq
\sum_{j'k'} \phi_{j'k'}^{(n)}\int\,N(s,\theta)  
\nabla_\perp(\Lambda_{j'}(s) \Lambda_{k'}(\theta)) 
    \nabla_\perp(\Lambda_j(s)
\Lambda_k(\theta)) \, J(s,\theta) \, \df s \, \df \theta= \frac{b_{jk}^{(n)}}{M^{(n)}},
\eeq
with: 
\aeq
N(s,\theta)=\sum_{s = {\mathrm{i}}} \frac{q_\mathrm{s}^2 n_{s}}{T_{s}} \rho_{s}^{2}.
\eeq
Note that this procedure relies on the polarization density factor $N(s,\theta)$ not depending on $\vphi$.
Practically, after the usual charge assignment, a DFT is applied on the spline coefficients in $\vphi$
to get the Fourier components $b_{jk}^{(n)}$. Each $b_{jk}^{(n)}$ is then divided by the analytically known term $M^{(n)}$,
which corresponds to the mass matrix of the B-spline basis in toroidal-Fourier space (see Ref.~\cite{Stier2024} for the
full derivation).
The resulting matrix problem is solved using a direct solver (LAPACK \cite{laug}).
Once the $\phi_{j'k'}^{(n)}$ Fourier coefficients are known, the electrostatic potential can be reconstructed
in any point in the domain.\\
Here we present a convergence order analysis of the Poisson solver using the Method of Manufactured Solutions (MMS) \cite{Oberkampf10}.
With this method, a manufactured solution is provided to the solver by adding proper source terms to the discrete equation. 
The deviation of the numerical solution of the equation compared to the provided manufactured solution is then checked.\\
In our specific case, we use the method of manufactured solution for the Poisson equation. The Poisson equation of PICLS has the form
\aeq\label{eq:mms}
N(f)[\phi_1]=S_{\phi_1}
\eeq
were $N(f)$ is a generic integro-partial differential operator.\\
We perform the following procedure:
\begin{enumerate}
\item First a specific source $S_{\phi_1}$ is constructed, for which a given function, $\phi_{1 MMS}$ is an exact solution of the system.
  The source $\phi_1$ is calculated by plugging the solution $\phi_{1 MMS}$ into equation \ref{eq:mms}. The function  $\phi_{1 MMS}$ is called the manufactured solution.
\item The normal solver routine is applied to get the solved potential $\phi_1$.
\item The constructed potential is then compared with the solved one to study the similarity of both and thus identify potential deviations.
  It is possible to calculate relative errors of the numerical solution and to verify the order of convergence of the numerical scheme. 
\end{enumerate}
\begin{figure}[h!tb]
\centering
\includegraphics[width=\textwidth]{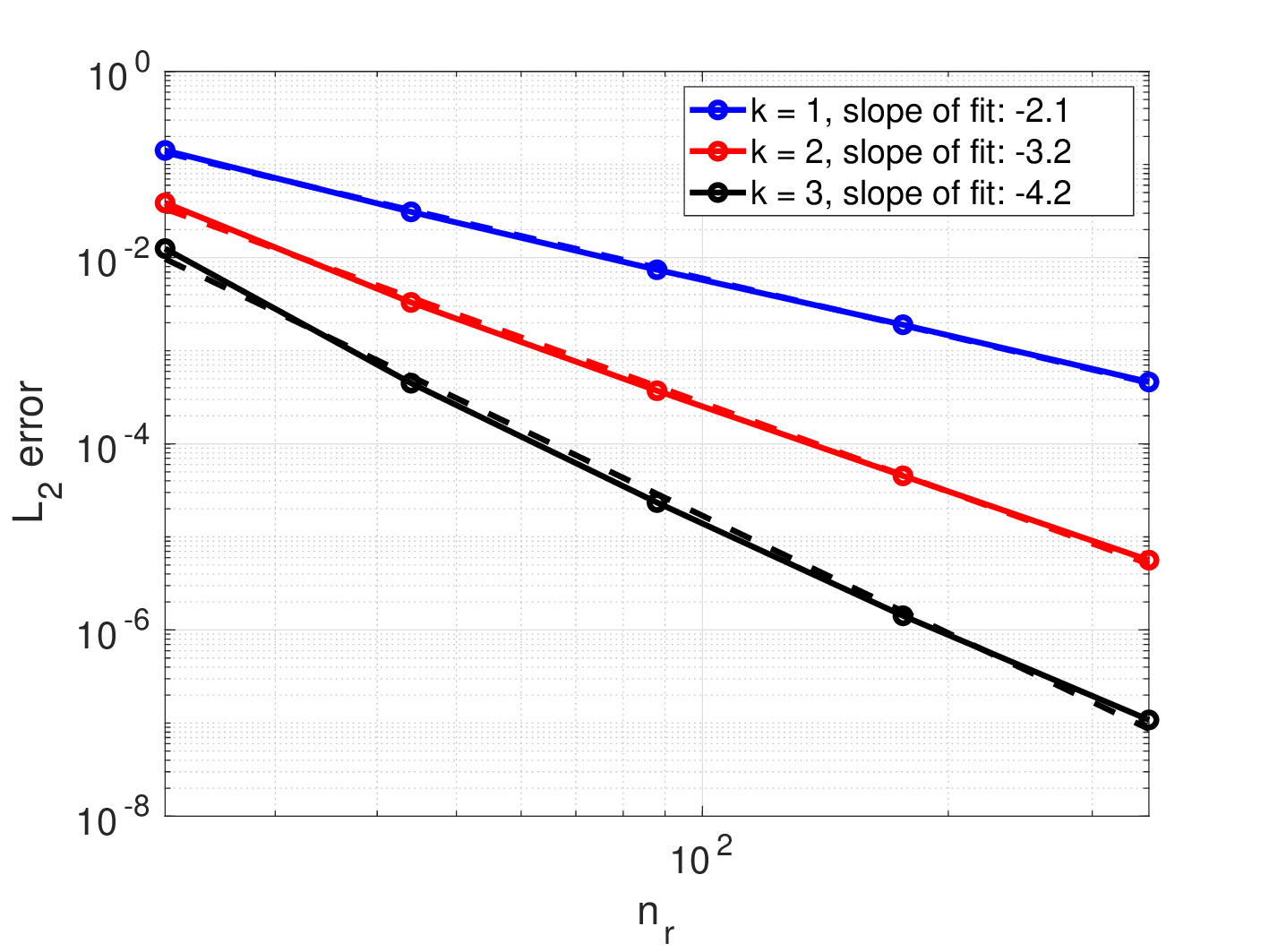}
\caption{Verification of the converge of the Poisson solver using the $L_2$ error. 
  The initial resolution is $(n_r=22,n_\theta=32,n_\varphi=8)$. Although the value of $n_r$ is used as label for the x-axis,
  the resolution in all directions was re-scaled by the same factor as $n_r$. Final resolution $(n_r=352,n_\theta=512,n_\varphi=128)$}
\label{fig:mms_test_l2}
\end{figure}
The chosen manufactured solution is
\aeq
\phi_{1 MMS}=\tilde{a}(g(r)+\alpha r+\beta)\sin(m_i\theta)\sin(n_i\varphi),
\eeq
with
\aeq
g(r)\equiv\exp\left(-\frac{1}{2}\left(\frac{r-r_0}{\sigma_r}\right)^2\right).
\eeq
Where $\tilde{a}$, $m_i$, $n_i$, $r_0$ and $\sigma_r$ are input parameters and $\alpha$ and $\beta$ are constants which are determined by the boundary conditions in $r$. Dirichlet boundary conditions are used, which imply $\beta=-g(r=0)$ and $\alpha=0$.
Assuming a flat density profile, and choosing $n_i=1$ and $m_s=1$, the corresponding source is: 
\begin{equation}\label{eq:mms_source}
        S_{\phi_1}  = -\frac{1}{r}\frac{1}{B^2}\frac{\partial \phi_{1 MMS}}{\partial r} + \frac{2}{B^3}\frac{\partial B}{\partial r}\frac{\partial \phi_{1 MMS}}{\partial r}-\frac{1}{B^2}\frac{\partial^2\phi_{1 MMS}}{\partial r^2}-\frac{1}{B^2r^2}\frac{\partial^2 \phi_{1 MMS}}{\partial \theta^2}
\end{equation}
In the following tests we used $m_i=5$, $n_i=2$, $q=2$, $r_0=3.02 \textrm{cm}$ and $\sigma_r = 0.6 \textrm{cm}$.
The screw-pinch geometry has the following parameters: $R=774 \textrm{cm}$, $B_0=1 \textrm{T}$ and $a= 6.05 \textrm{cm}$. The simulation box ranges from $r=0$ to $r=6.05 \textrm{cm}$. As mentioned before, the density profile is flat. The initial resolution is determined by the number of splines in the three directions, set to $(n_r=22, n_\theta= 16, n_\phi=8)$. The convergence of the algorithm is tested by doubling the resolution in each direction. Given a discrete vector field $\phi_p=\phi(r_i,\theta_j,\phi_k)$ with $i=[1:\bar{n}_r]$, $j=[1:\bar{n}_\theta]$ and $k=[1:\bar{n}_\varphi]$, we define an error, $L_2$, using the $L_2$ norm 
\aeq
\left\Vert \phi \right\Vert_{L_2}=\sqrt{\Sigma_p{\phi_p^2}}, 
\eeq
with
\aeq
L_2=\frac{\left\Vert \phi - \phi_{MMS}\right\Vert_{L_2}}{\left\Vert\phi_{MMS}\right\Vert_{L_2}}
\eeq
Results are shown in Fig.~\ref{fig:mms_test_l2} for linear ($k=1$), quadratic ($k=2$) and cubic ($k=3$) splines. While only $n_r$ is reported as a label for the x-axis, actually the number of splines in all directions has been re-scaled, ranging from $(n_r=22,n_\theta=32,n_\varphi=8)$ to $(n_r=352,n_\theta=512,n_\varphi=128)$. The slope of the linear fit on the log-log plot is reported in the legend of the two figures.  The slope of the error curve is mathematically expected to be equal to $k+1$ which matches the results shown here in good approximation. In those tests the number of grid points used for comparing the two solutions are $\bar{n}_r=90$, $\bar{n}_\theta=100$ and $\bar{n}_\varphi=110$.\\

\subsection{Geometries}
\begin{figure}[htb]
\centering
\includegraphics[width=\textwidth]{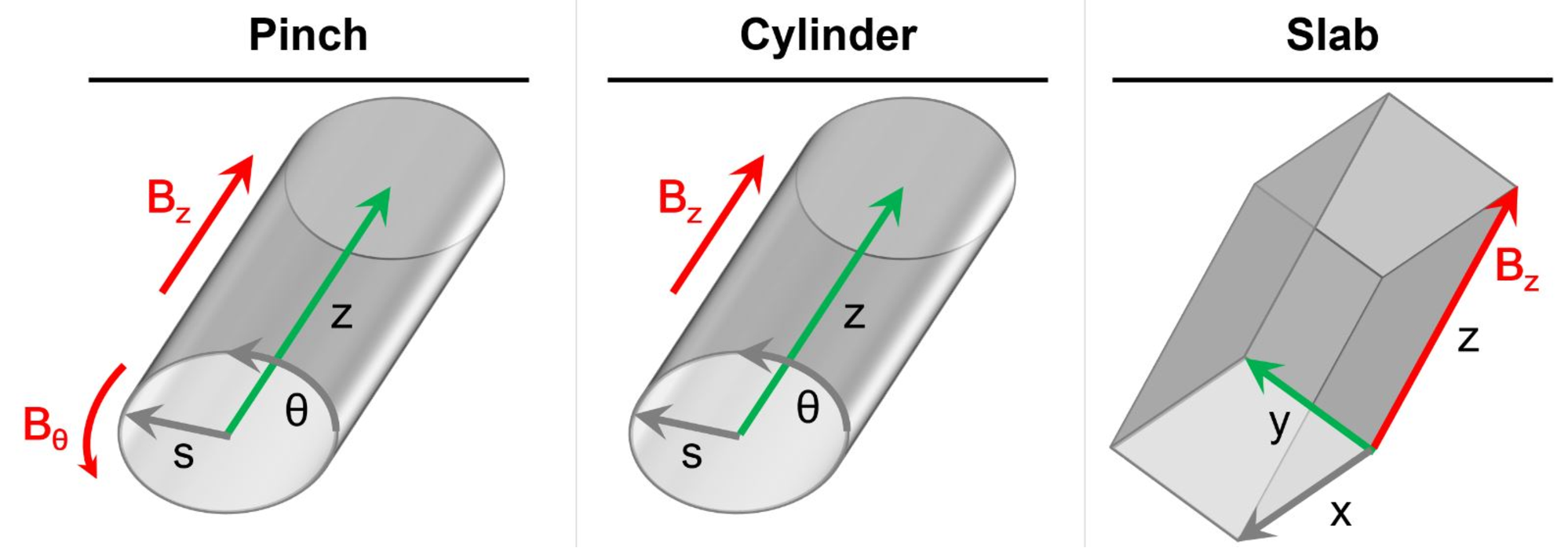}
\caption{Illustration of the three different geometries ``Pinch'', ``Cylinder'', and ``Slab'' that are currently implemented in PICLS. ``Pinch'' and ``Cylinder'' are implemented in cylindrical coordinates $(s,\theta,z)$, whereas ``Slab'' uses $(x,y,z)$ coordinates. The red arrows indicate the direction of the B-field of the respective geometry. The green arrows indicate the direction of the coordinate for the DFT of the electric field solver.
}
\label{geometries_fig}
\end{figure}

In PICLS the markers are loaded in a Cartesian space of dimension $(L_x,L_y,L_z)$. The particle motion in the absence of fluctuations is determined by
the background magnetic field. Note that an external electrostatic potential can also be imposed on the particle equations of motion. Moreover, two different elliptic
solver are implemented in PICLS. The first one is in Cartesian coordinates, where the $y$ coordinate is periodic while (zero or nonzero) Dirichlet boundary conditions can be applied in $x$ and $z$. The second one is similar to the ORB5 solver, based on polar coordinates, in which only one coordinate (radius) is non periodic.
The main difference between the two solver types is that in the polar case the FT-direction lies along the main B-field direction while in the Cartesian solver
it is directed perpendicularly to the B-field. 
At the moment, three different magnetic geometries are implemented in PICLS (see also Fig.~\ref{geometries_fig} for illustration): 
\begin{enumerate}
	\item \textbf{Cylinder:} In this case, the geometry is cylindrical with the coordinates $(s,\theta,z)$, where $s$ is the radius $r$ of the cylinder normalized to the minor radius $a$, $\theta$ the poloidal angle and $z$ goes in the ``toroidal'' direction. The cylinder geometry can also be used as cylindrical Tokamak, for which $z$ is replaced by the toroidal angle $\vphi$. In this case, the magnetic field $B$ has only a component in $z$ direction and also the Fourier transformed direction is along $z$. In this case, only one direction, $s$, can be non periodic.
	\item \textbf{Pinch:} This geometry in general is similar to the ``Cylinder'' case, except that the B-field can have an additional component in the poloidal $\theta$ direction.
	\item \textbf{Slab}: Here, we use a 3D slab geometry, with the spatial
          coordinates $(x,y,z)$. The magnetic field is aligned along the $z$
          coordinate, but different from the ``Cylinder'' and ``Pinch'' case,
          the Fourier transformed direction is along the perpendicular $y$
          direction. Both $x$ and $z$ direction are non periodic and on $z$
          (logical) sheath boundary conditions can be applied.
          The logical sheath model used PICLS is based on the Parker's model
          (\cite{parker93sheath}). The implementation of this
          model in PICLS can be found, for the 1D case,  in
          \cite{Boesl2019}. In the 3D code, the total parallel current to the
          wall is set to zero ($j_\parallel=0$) at every point in time and
          thus the wall can be regarded as insulating. From a physical point
          of view, incident ions that flow towards the
          sheath boundary are accelerated by the dropping sheath potential.
          Whereas electrons are only absorbed if
          their velocity is high enough to overcome the sheath potential drop
          at the wall.
          Electrons below this critical velocity are reflected back into the
          domain. Details and validation efforts for the 3D PICLS sheath model will be
          published elsewhere. In the future, more sophisticated sheath models could be considered (see, e.g.~\cite{TOGO2016109}).

\end{enumerate}
\subsection{Parallelization} \label{part_para_sec}
\begin{figure}[htb]
\centering  
\includegraphics[width=\textwidth]{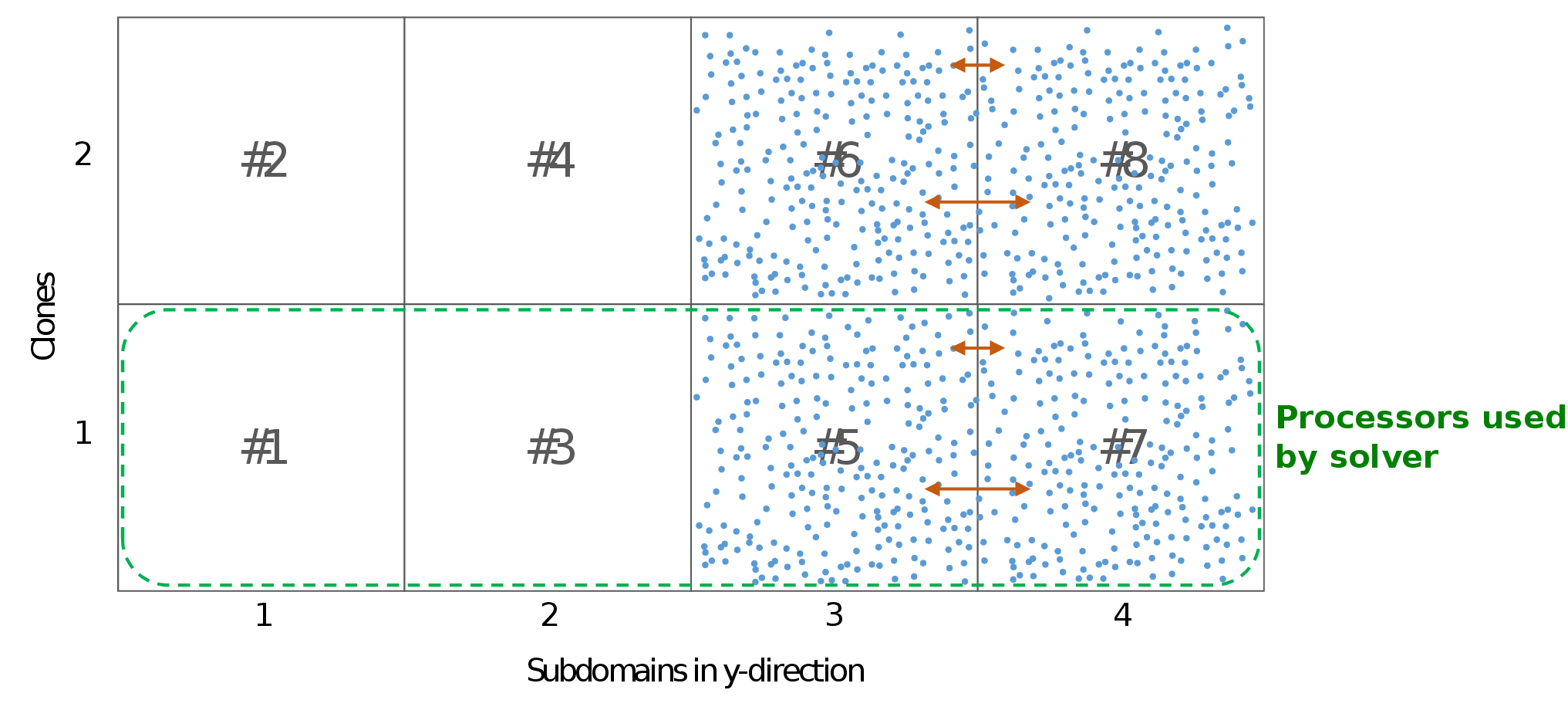}
\caption{Illustration of the PICLS parallelization scheme with 4 domains and two clones. The blue dots represent the particles and the orange arrows the communication between the MPI domains. In the example, the communication is displayed exemplary for domains 3 \& 4 and clones 1 \& 2. With the green box we indicated to which processors the particle sums are reduced for the field solve. The total number of processors (indicated by \#) is equal to the number of domains times the number of clones.
}
\label{parallel_scheme}
\end{figure}

Running high resolution plasma simulations with a gyrokinetic full-f PIC code, in general means that a significant number of markers and grid cells is required. 
However, the more markers and cells we introduce, the higher the computational costs become. 
As a result, an efficient parallelization scheme needs to be implemented, to share data between multiple processors to join their computation resources and to optimize the memory use.\\
In PICLS, we use a hybrid OpenMP and MPI approach. The specific methods we apply for the parallelization of the MPI tasks are called domain decomposition and domain cloning.\\
In the domain decomposition approach, the grid cells are split up into different domains. Each domain is then attributed to a MPI task and also the particles that are present within the cell at the specific time-step are attributed to this task. For our domain decomposition, we split the cells along the periodic Fourier transformed direction.\\
In the domain cloning \cite{HATZKY2006325} approach, however, the number of particles is divided between the number of defined MPI clones. Each clone has a copy of the whole domain field data to deposit the charges. The sums of the charges of each clone are then added up via MPI communication to calculate the source term for field equation. The fields are then solved and broadcast to all clones.\\
Each MPI task eventually is responsible for a specific clone and a specific domain. Within each of these MPI tasks, the particle operations are performed in parallel and OpenMP is applied, to optimally use computational resources. Since the particles can move from one domain to the other, according to their characteristic equations, a function "particle\_move" is implemented that moves the particles from the exiting to the entering MPI domain for the calculations. 
In figure \ref{parallel_scheme} this parallelization scheme is sketched.\\
Preliminary performance results on the MARCONI FUSION HPC system at CINECA
showed a speedup of 2.8 for realistic ITG turbulence simulations (see next
Section) when moving from 20 nodes (960 cores) to 80 nodes (3840 cores) using
128 million markers. Moreover, the weak scaling on number of marker is
practically perfect when moving from 20 to 80 nodes.\\ 
Apart from general code optimization, additional effort will be needed to port the code to GPUs to fully exploit the capabilities of new computing systems. Potential options here are OpenACC and OpenMP-offload, which have to be further evaluated.

\section{Verification}
In this section we present some verification efforts for closed-field-line configurations, including a comparison with two production codes: the grid based
GENE-X \cite{Michels20} code and the PIC code ORB5 \cite{lantiCPC2020}. All the simulations discussed in this section were performed using cubic B-splines ($k=3$).
\subsection{ITG modes,comparison with GENE-X}
Going from a 1D to a 3D spatial domain is not trivial and requires careful implementation and testing. Several main aspects of our
PIC algorithm, especially the field solver and the particle pusher, needed to be adjusted to allow for higher dimensional simulations.
Therefore, the new code has been verified on the known problem of ITG (Ion Temperature Gradient) instabilities in a screw-pinch geometry, for
which pre-existing results exist \cite{Michels20}.
\begin{figure}[htb]
\centering  
\includegraphics[width=120mm]{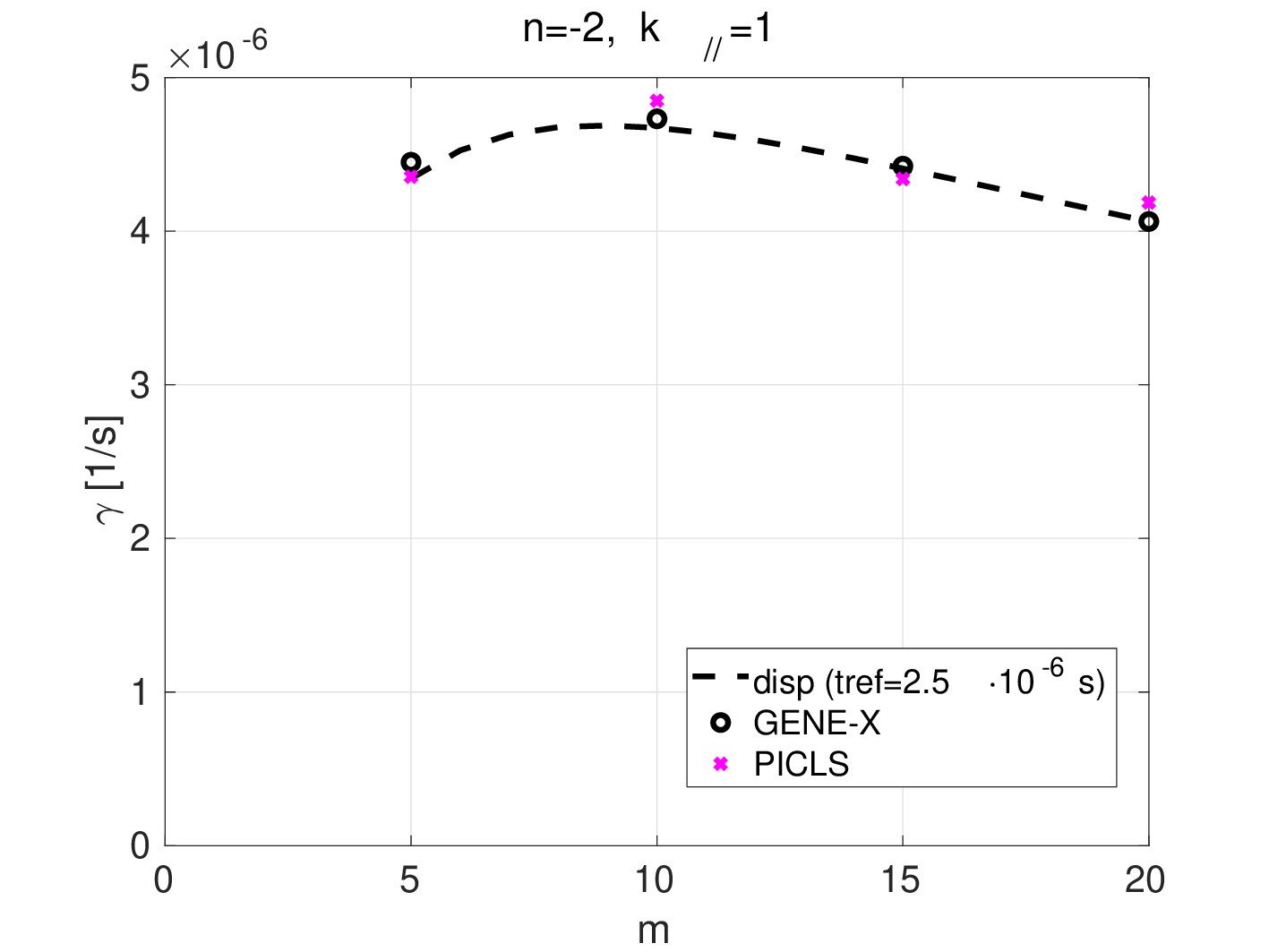}
\caption{Growth rates measured via linear fits on a logarithmic plot of the $L_2$ norm of the potential, for different
poloidal mode numbers, for a fixed $n=-2$ and for q-values adjusted to keep
$k_\parallel=1$ constant. PICLS results are in good agreement with both GENE-X
and dispersion relation data.}
\label{fig:genexcomp}
\end{figure} 

\begin{figure}[htb]
\centering  
\includegraphics[width=120mm]{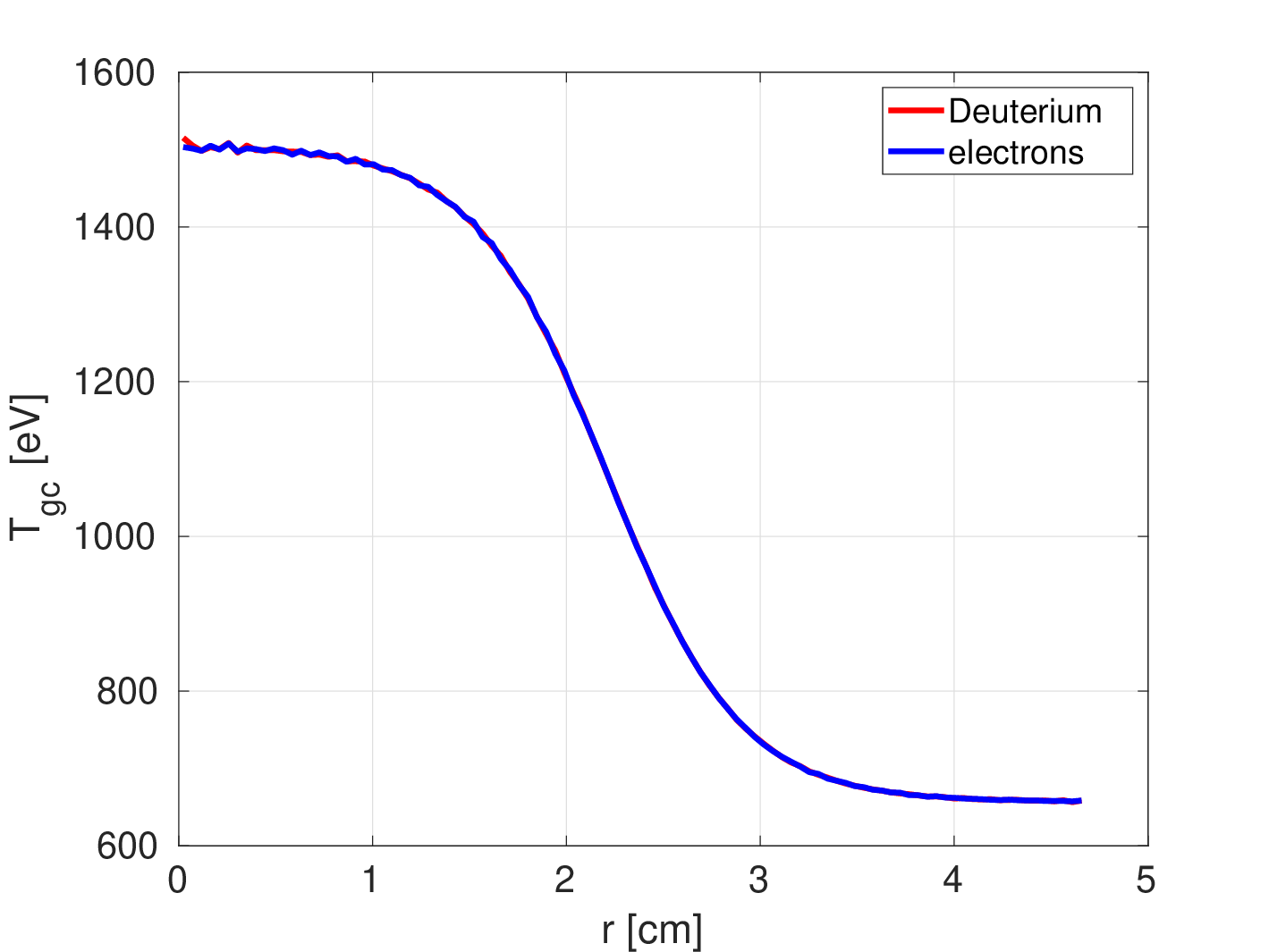}
\caption{Temperature profiles calculated as moment of the marker distribution function, showing that the newly implemented Maxwellian loading
routine provides the expected behavior in velocity space}
\label{fig:genexprofile}
\end{figure}

The screw-pinch configuration (a.k.a.~{\em straight Tokamak}) is a periodic system, based on a cylindrical pinch device with the coordinates $(r, \theta, z)$, where $r$ is the radius,
$\theta$ the poloidal angle and z the coordinate along the cylindrical axis (or toroidal direction). 
Its name comes from the helical shape of the magnetic field with components in the $z$ and $\theta$ direction. 
The electric potential solver we use for the screw-pinch is the solver in polar coordinates, with an DFT in the periodic toroidal $z$-direction, periodic boundary conditions in the poloidal direction and Dirichlet boundary conditions in the radial direction. 
The particles leaving the domain in the radial direction are destroyed and recreated randomly into the physical domain in order to keep the total number
of physical particles fixed. The screw-pinch magnetic field in polar coordinates is
\aeq \label{eq:vecB_pinch}
\vecb{B} = B_0 \left(-\frac{r}{Rq_\textrm{s}(r)}\btheta+\vecb{z}\right),
\eeq
with the radius $r$, the q-factor $q_\textrm{s}(r)=q_0+q_1 s^2$, the normalized radius $s=r/a$, the constant magnetic field $B_0$ and $R=L_z/2\pi$ ($L_z$ is the length of the cylinder in z-direction). For a screw-pinch, Maxwellians are an exact equilibrium solution of the gyrokinetic system of equations. Therefore, in order to destabilize ITG modes
an initial perturbation with a Gaussian radial shape and a fixed toroidal and poloidal number component is added to the ion density.
Moreover, according to the dispersion relation, the Eigenvalues of a slab ITG mode are functions of the poloidal mode number $m$ and the parallel wave number $k_\parallel$.\\
The simulation setup is described in \cite{Michels20} and summarized in Table~\ref{tab3}.

\begin{table}
  \caption{\label{tab3}PICLS geometric setup and initial perturbation for the four simulations of Fig.~\ref{fig:genexcomp}. Each of the four
  simulations is characterized by different values of $q$. The poloidal and toroidal mode numbers are selected such that $k_\parallel=1$
  in all cases. Consequently, in each simulations the value of $m$ has been adapted to the corresponding $q$:  $m=5$ for $q=5/3$,
  $m=10$ for $q=10/3$, $m=15$ for $q=15/3$ and $m=20$ for $q=20/3$.}
   \begin{indented}
   \item[]\begin{tabular}{@{}lll}
     \br
     Parameter & Value & Description \\
     \mr
     a & 4.68 [\textrm{cm}] & minor radius\\
     R & 77.0 [\textrm{cm}] & major radius\\
     $n_0$ & $0.5\cdot 10^{13}~[\textrm{cm}^{-3}]$ & density flat\\
     $T_{e0}=T_{i0}$ & $1000~[\textrm{eV}]$ & Temperature at $s=0.5$\\
     $m_e$ & $1~m_{\textrm elec}$ & electron mass\\
     $m_i$ & $3972~m_{\mathrm elec}$ & ion mass (D)\\
     $\rho_s$ & 0.32 [\textrm{cm}] & sound Larmor radius\\
     $k_\parallel$ & 1 & parallel wave number (fixed)\\
     $n$ & -2 & toroidal mode number\\ 
     $q$ & $5/3$, $10/3$, $15/3$, $20/3$ & safety factors\\  
     $m$ & 5, 10, 15, 20 & poloidal mode numbers\\
  \end{tabular}
\end{indented}
\end{table}

Note that in all those simulations a single $n$ mode, $n=-2$ has been included in the simulation.
The background temperature has been defined following the prescription provided in \cite{Michels20}.
The results, summarized in Fig.~\ref{fig:genexcomp}, show an excellent agreement with the GENE-X results (from \cite{Michels20})
and the dispersion relation.\\
In PICLS, the growth rate has been calculated by taking a linear fit on the logarithmic plot of the $L_2$ norm of the electrostatic potential,
during the linear phase of the simulation.
Fig.~\ref{fig:genexprofile} shows the ion and electron temperature profiles reconstructed by taking moments
of the marker distribution function. This shows that the newly introduced importance sampling routines provide
the expected Maxwellian loading in velocity space.

\subsection{ITG turbulence}
\begin{figure}[htb]
\centering  
\includegraphics[width=\textwidth]{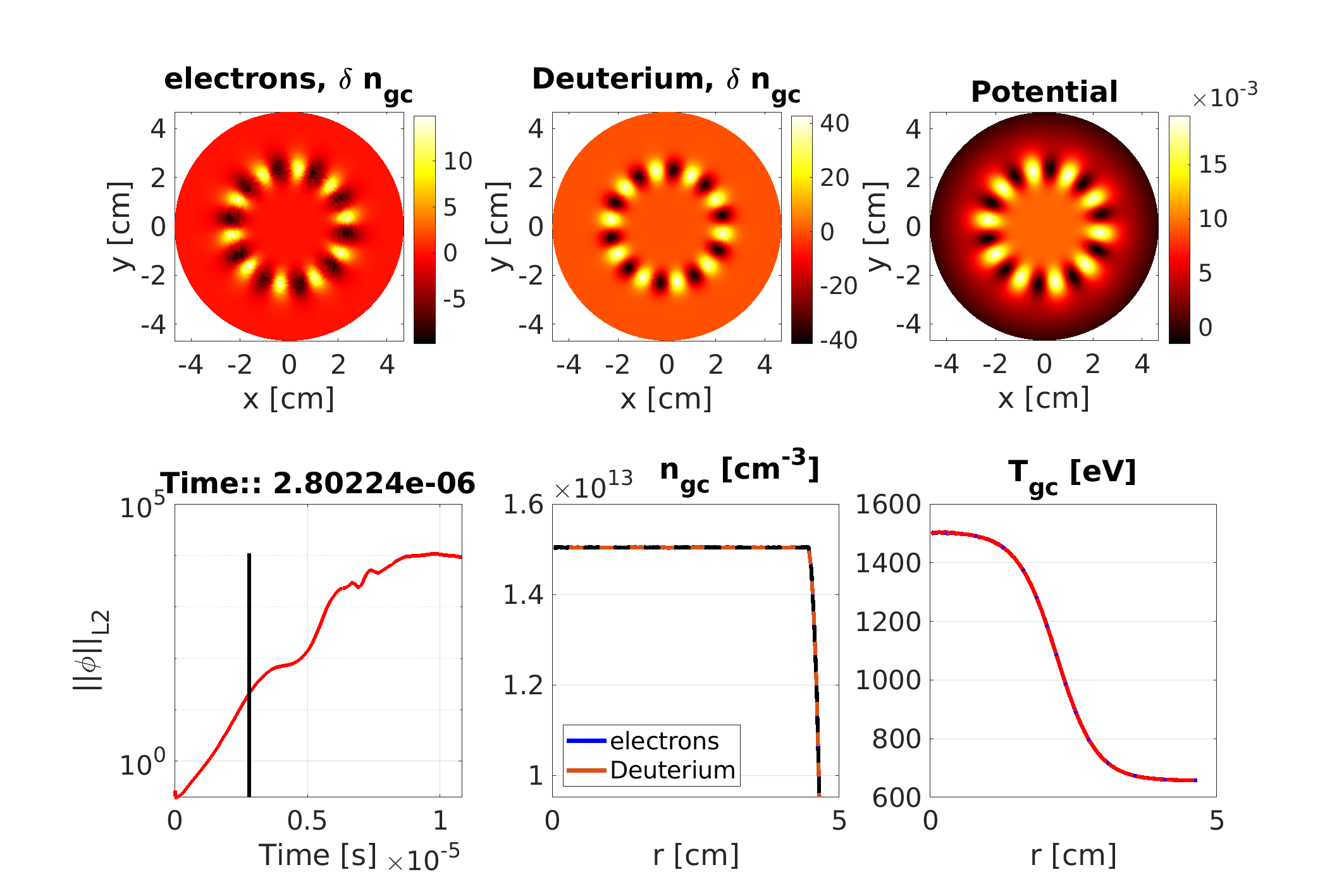}
\caption{Top line: snapshot of the poloidal cross section of the electron density fluctuation (left), ion gyrocenter density fluctuation (middle) and
  electrostatic potential (right) at the end of the linear phase. Bottom line: time evolution of the $L_2$ norm of the electrostatic potential (left) and snapshot of the
  density (middle) and temperature (left) at the end of the linear phase. The
  resulting electrostatic potential correctly reflects the properties of the
  linear elliptic field equation solved by PICLS. 
}
\label{fig:turbulence3}
\end{figure}
\begin{figure}[htb]
\centering
\includegraphics[width=\textwidth]{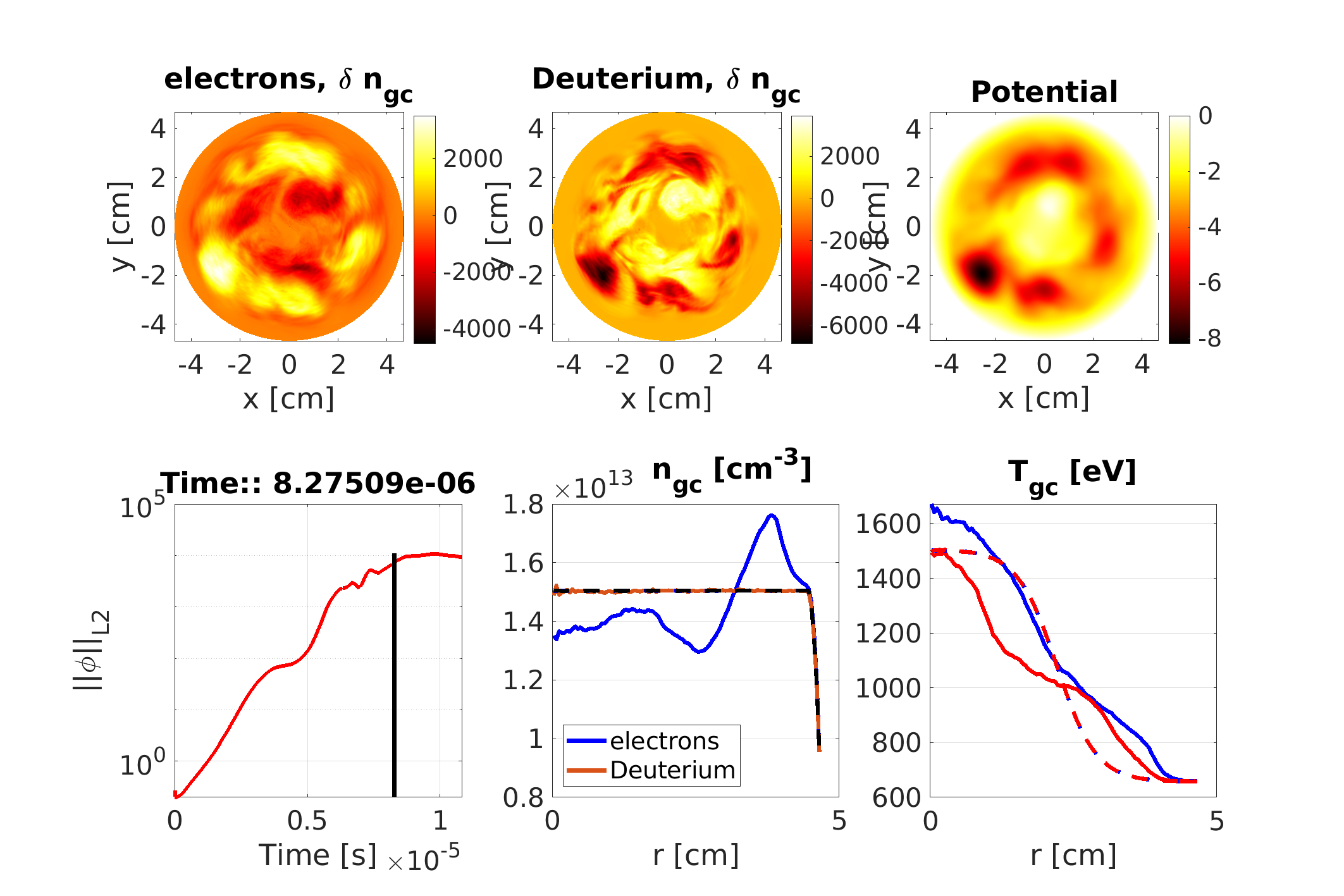}  
\caption{Top line: snapshot of the poloidal cross section of the electron density fluctuation (left), ion gyrocenter density fluctuation (middle) and
  electrostatic potential (right) during the saturation phase. Bottom line: time evolution of the $L_2$ norm of the electrostatic potential (left) and snapshot of the
  density (middle) and temperature (left) during the saturation phase. 
}
\label{fig:turbulence4}
\end{figure}
Using the same numerical setup of the GENE-X benchmark, described in the previous section, a set of turbulence simulations has been performed
by increasing the number of toroidal mode numbers allowed in the simulation. Fig.~\ref{fig:turbulence3} shows an example of such a simulation
in which six toroidal mode numbers, $n=[0:5]$ are retained, for the $q=10/3$ case. The initial perturbation has been
kept the same as in the $k_\parallel=1$ and $n=-2$ case previously described. The relevant numerical parameters are described in Table~\ref{tab4}.
In all the multi-n simulations discussed in this section a Maxwellian control variate has been used, in order to reduce the statistical noise.

\begin{table}
  \caption{\label{tab4}PICLS numerical setup for ITG turbulence simulations.}
   \begin{indented}
   \item[]\begin{tabular}{@{}lll}
     \br
     Parameter & Value & Description \\
     \mr
     nptot el. & $128~{\textrm M}$ & number of electron markers\\
     nptot ions & $128~{\textrm M}$ & number of ion markers\\
     $n_r$ & 100 & number of splines in $r$\\
     $n_{\theta}$ &  128& number of splines in $\theta$\\
     $n_{\vphi}$ & 32  & number of splines in $\vphi$\\
     dt & $2.\cdot 10^{-8}~{\textrm s}$& time step\\
  \end{tabular}
\end{indented}
\end{table}

The top three panels of Fig.~\ref{fig:turbulence3} show a snapshot of the poloidal cross section of the electron density (left), the ion gyrocenter density (middle)
and the electrostatic potential (right), toward the end of the linear phase.
During the linear phase temperature and density profiles do not evolve in time, as it is clearly illustrated by the bottom middle and right panels
of Fig.~\ref{fig:turbulence3}.

The time evolution of the $L_2$ norm of the electrostatic potential is shown in the left panel of the bottom line. The vertical black line indicates
the time at which the snapshot of the other five panels was taken. We can identify three phases characterizing the time evolution of the system.
During the first phase, the initial perturbation grows linearly in time, being an unstable Eigenmode of the system. At around time $T=4\cdot10^{-6}~\textrm{s}$ a second
mode, with higher linear growth rate (and lower $k_\parallel$) becomes the most unstable mode and dominates the dynamic.

\begin{figure}[htb]
\centering  
\includegraphics[width=\textwidth]{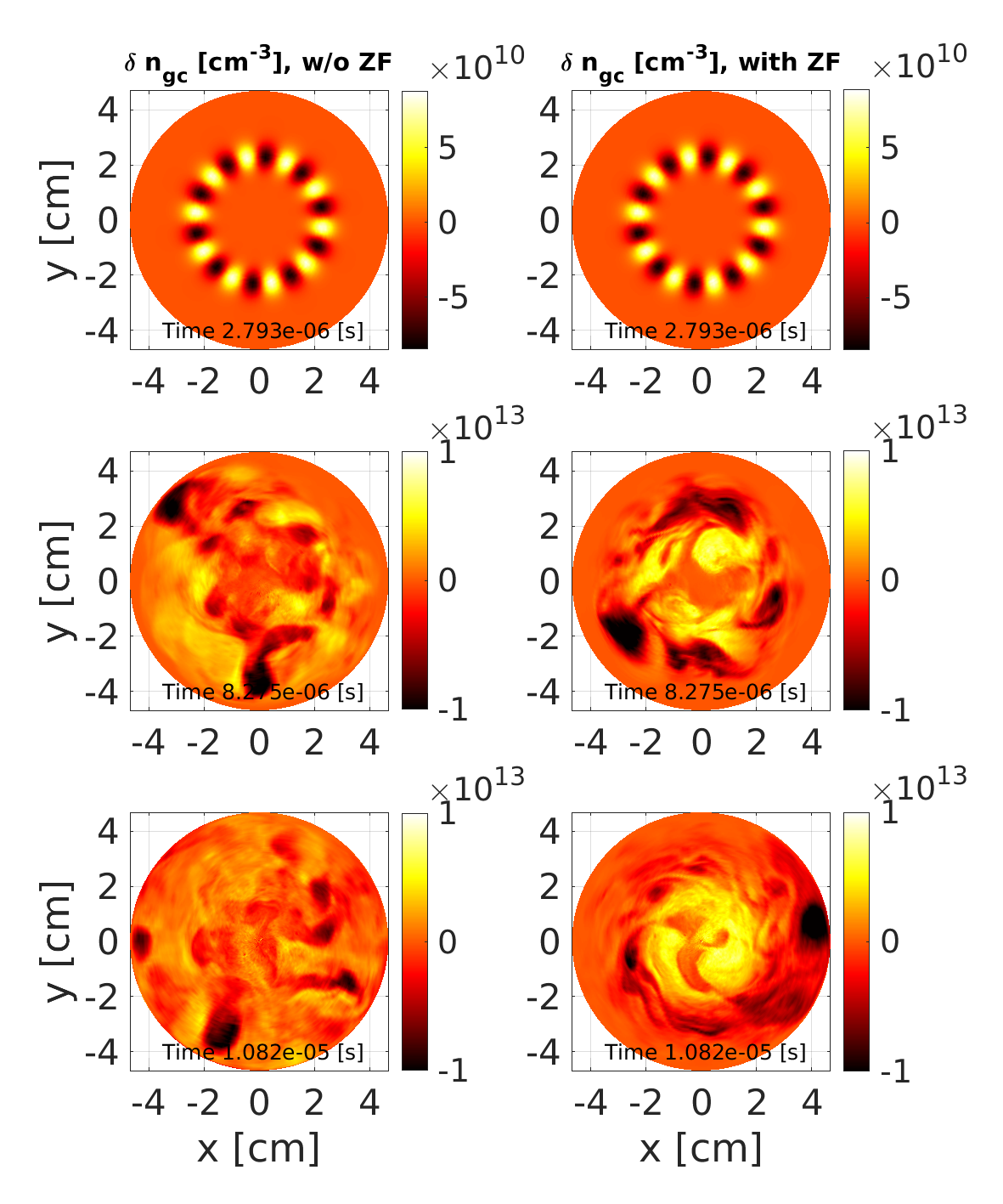}  
\caption{Snapshots of the poloidal cross section of the ion gyrocenter density
  fluctuations during the linear phase (top), the early 
  saturation phase (middle) and at the last time step (bottom) for two
  simulations w/o (left) and with (right) Zonal Flows. The Zonal Flow has been
  suppressed by filtering out the $n=0$ component in the field solver.
  During the saturation phase, large density blobs
  are created, with an amplitude even larger then the initial background density.} 
\label{fig:turbulence5}
\end{figure}

Figure \ref{fig:turbulence4} illustrates what happens
at later times, when nonlinear terms dominate the dynamics and the turbulence saturates. The electrostatic potential clearly shows the formation
of zonal structures, the so-called zonal flows (ZFs). Note that ZFs are not the only possible saturation mechanism in this kind of simulations since no sources are applied
to prevent the relaxation of the temperature profiles and the consequently reduced drive. Temperature profile relaxation is indeed visible in both the ion and electron temperatures of Fig.~\ref{fig:turbulence4}.
Note that also the boundary conditions can play a significant role in determining the saturation level of such a system. \\
Nevertheless, it is important to verify that PICLS is able to catch the physics related to ZFs. Therefore, we have repeated the same
simulation but keeping only $n=[1:5]$ modes in the solver, thus suppressing the zonal component of the potential. 
Figure ~\ref{fig:turbulence5} compares the ion density fluctuations of the two
simulations w/o (left) and with (right) ZFs, during the linear phase (top),
the early saturation (middle) and at the end of the simulation (bottom). Being the ZFs linearly stable,
they do not play any role during the linear evolution of the system. On the other hand, at saturation, density blobs tend to rotate along the zonal potential
structures, while they drift rapidly toward the plasma edge when the ZF is filtered out.
Moreover, Fig.~\ref{fig:turbulence5} also illustrates the capability of PICLS of dealing with large fluctuations, of the order
of the background density.

\subsection{Omega-H mode, comparison with ORB5}
The case discussed in this section is inspired by Ref.~\cite{Biancalani2016} where
a Shear \Alfv~Wave (SAW) was simulated using ORB5. This test case was specifically
designed to minimize the kinetic corrections to the wave. This was  achieved by
choosing a Tokamak magnetic equilibrium with very large aspect
ratio ($R/a=100$) and a flat q profile. An initial ion density perturbation was imposed, with a radial Gaussian shape and a
poloidal mode number $m=1$ ($n=0$). This initial  perturbation also excites a damped sound wave, making the interpretation of the
results complicated (two damped waves in the system). Therefore, the ion weight distribution was reset to the Maxwellian distribution
after the first call to the solver, in order to suppress the sound wave related to the initial ion pressure perturbation.\\
As mentioned before, PICLS is an electrostatic code, although an electromagnetic version is under development. Therefore, a direct
comparison with the published ORB5 results is not possible.
Nevertheless, the SAW has an electrostatic counterpart, the so-called Omega-H mode. The Omega-H mode
\cite{Lee83} is an Eigenmode of the electrostatic system which does not have a correspondent in nature.
The dispersion relation of the Omega-H mode is obtained by taking the $\beta\rightarrow 0$ limit while keeping
the sound velocity $c_s$ finite in the kinetic dispersion relation of a SAW.
The simplest SAW kinetic dispersion relation, see e.g. \cite{Lysack23} is 
\aeq
\omega_{SAW,KIN}^2 = v_A^2 \frac{1+k_\perp^2\rho_s^2}{1+k_\perp^2d_e^2}k_\parallel = 2c_s^2 \frac{1+k_\perp^2\rho_s^2}{\beta_e+k_\perp^2\rho_s^2}k_\parallel.
\eeq
The Omega-H dispersion relation is obtained by setting $\beta_e=0$ but keeping $c_s$ finite, leading to 
\aeq
\omega_H^2 = 2c_s^2 \frac{1+k_\perp^2\rho_s^2}{k_\perp^2\rho_s^2}k_\parallel,
\eeq
with
\aeq
c_s^2=\frac{T_e}{m_1},v_A^2=\frac{B_0^2}{\mu_0nm_i},~~\beta_e=\frac{2\mu_0p_e}{B_0^2}=2\frac{c_s^2}{v_a^2},\nonumber
\eeq
\aeq
d_e^2=\frac{m_i}{2\mu_0nq_i},~~\rho_s=\frac{c_s}{\Omega_{ci}},~~\Omega_{ci}=\frac{q_iB_0}{m},~~k_\parallel=\frac{1}{qR}\nonumber.
\eeq

\begin{figure}[htb]
\centering  
\includegraphics[width=120mm]{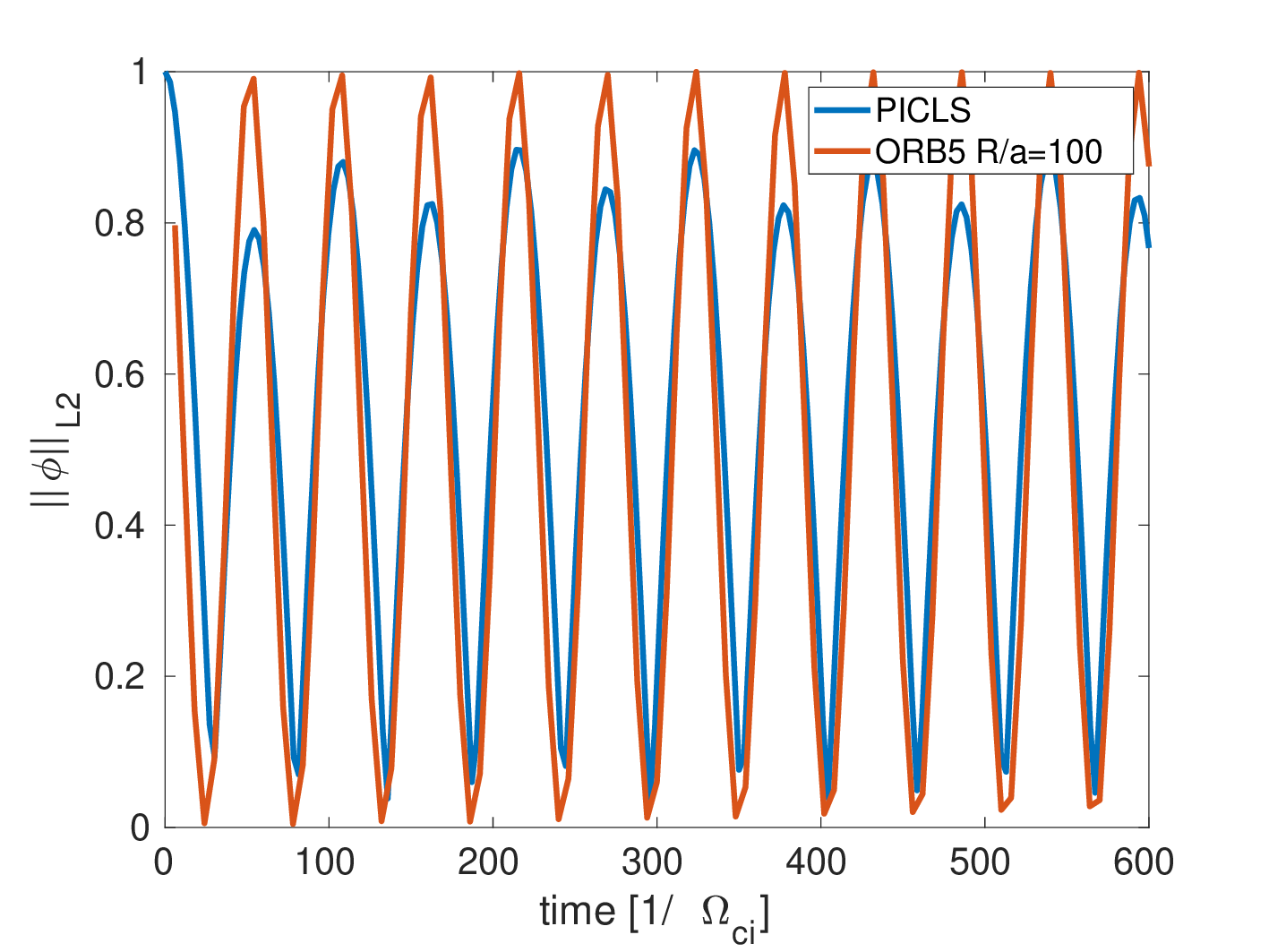}  
\caption{Time evolution of the $L_2$ norm of the electrostatic potential for ORB5 and PICLS, using the setup described in Table~\ref{tab5}.}
\label{fig:orb5comp}
\end{figure} 

\begin{figure}[htb]
  \centering
  \includegraphics[width=120mm]{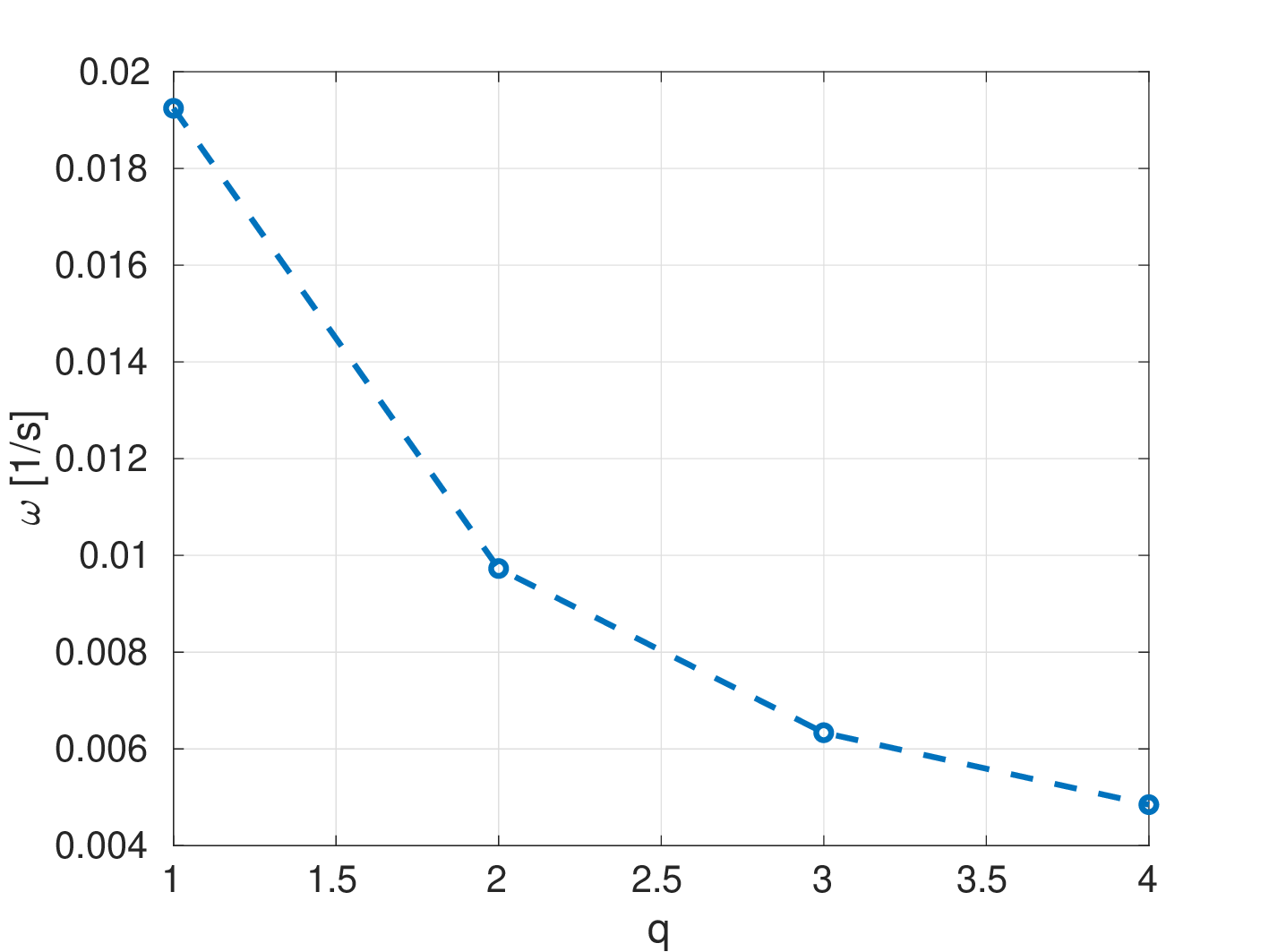}  
  \caption{Frequency of the Omega-H mode as a function of the (constant) safety factor using PICLS with the setup described in Table~\ref{tab5}.}
\label{fig:qscan}
\end{figure} 

The code ORB5 has been rerun using the same input of Ref.~\cite{Biancalani2016}, with the exception of the choice of the electrostatic
model for the field equations. Fig.~\ref{fig:orb5comp} shows the time evolution of the $L_2$ norm (in arbitrary units) for the two codes, showing
an excellent agreement on the frequency of the Omega-H mode. Note that the background temperature has been set to a very low value ($100~{\textrm eV})$ in order
to minimize the Landau damping. Fig.~\ref{fig:qscan} shows the dependency of the frequency of the Omega-H mode on the safety factor q in PICLS simulations.
The expected $1/q$ dependence in $k_\parallel$ of the dispersion relation is retrieved.

\begin{table}
  \caption{\label{tab5}PICLS geometric setup and initial perturbation for the simulations of Figs.~\ref{fig:orb5comp} and \ref{fig:qscan}.}
   \begin{indented}
   \item[]\begin{tabular}{@{}lll}
     \br
     Parameter & Value & Description \\
     \mr
     a & 1.67 [\textrm{cm}] & minor radius\\
     R & 167 [\textrm{cm}] & major radius \\
     $n_0$ & $10^{15}~[\textrm{cm}^{-3}]$ & density (flat)\\
     $B_0$ & $2.4 [\textrm T]$ & magnetic field on axis\\
     $T_{e0}=T_{i0}$ & $100~[\textrm{eV}]$ & temperature(flat)\\
     $m_{init}$ & 1 & initial poloidal mode \\
     $n_{init}$ & 0 & initial toroidal mode \\
     $q$ & 2 (flat) & safety factor`\\
     $m_e$ & $1~m_{\textrm elec}$ & electron mass\\
     $m_i$ & $3972~m_{\mathrm elec}$ & ion mass\\
  \end{tabular}
\end{indented}
\end{table}

\section{Conclusions}
The PICLS code has been extended to a 3D spatial domain. Since going from 1D to 3D included significant changes in the models and algorithms applied, we
have presented here a number of verification tests for closed-field-line geometry. Key features of the 3D code are 
the 3D importance sampling particle loading, the gyroaverage algorithm and the Fourier enhanced field solver \cite{Stier2024}.
Concerning the field solver, we have applied the Method of Manufactured Solutions
to a physically relevant 3D case. Finally, we have verified the code by comparing with the two well established GENE-X and ORB5 codes, for two different problems.
Both verification tests were passed. We have also shown that PICLS can cope with large density fluctuations, comparable in size or even large than the background,
which is a crucial requirement for future SOL applications.\\

\ack{}
The authors wish to thank Roman Hatzky and Emanuele Poli for fruitful discussions.\\
This work has been carried out within the framework of the EUROfusion Consortium, partially funded by the European Union via the Euratom Research and Training Programme (Grant Agreement No 101052200 — EUROfusion). The Swiss contribution to this work has been funded by the Swiss State Secretariat for Education, Research and Innovation (SERI). Views and opinions expressed are however those of the author(s) only and do not necessarily reflect those of the European Union, the European Commission or SERI. Neither the European Union nor the European Commission nor SERI can be held responsible for them. Part of the simulations discussed in this work were performed on the MARCONI FUSION HPC system at CINECA. \\

\bibliography{literature}

\begin{thebibliography}{10}

\bibitem{Zweben07}
S.~J. Zweben, J.~A. Boedo, O.~Grulke, C.~Hidalgo, B.~LaBombard, R.~J. Maqueda,
  P.~Scarin, and J.~L. Terry.
\newblock {Edge turbulence measurements in toroidal fusion devices}.
\newblock {\em {PLASMA PHYSICS AND CONTROLLED FUSION}}, {49}({7}):{S1--S23}, 07
  2007.
\newblock {IEA Large Tokamak IA Workshop on Edge Transport in Fusion Plasmas,
  Cracow, POLAND, SEP 11-13, 2006}.

\bibitem{stangeby2000plasma}
P.C. Stangeby.
\newblock {\em The plasma boundary of magnetic fusion devices}.
\newblock CRC Press, 2000.

\bibitem{Ricci15}
P.~Ricci.
\newblock Simulation of the scrape-off layer region of tokamak devices.
\newblock {\em Journal of Plasma Physics}, 81(2):435810202, 2015.

\bibitem{Mosetto13}
A.~Mosetto, F.~D. Halpern, S.~Jolliet, J.~Loizu, and P.~Ricci.
\newblock Turbulent regimes in the tokamak scrape-off layer.
\newblock {\em Physics of Plasmas}, 20(9):092308, 2013.

\bibitem{Cohen94}
R.~H. Cohen, N.~Mattor, and X.~Xu.
\newblock Scrape-off layer turbulence theory.
\newblock {\em Contributions to Plasma Physics}, 34(2-3):232--246, 1994.

\bibitem{ISOARDI20102220}
L.~Isoardi, G.~Chiavassa, G.~Ciraolo, P.~Haldenwang, E.~Serre, Ph. Ghendrih,
  Y.~Sarazin, F.~Schwander, and P.~Tamain.
\newblock Penalization modeling of a limiter in the tokamak edge plasma.
\newblock {\em Journal of Computational Physics}, 229(6):2220--2235, 2010.

\bibitem{Tskhakaya09}
David Tskhakaya, R.A. Pitts, Wojciech Fundamenski, T~Eich, S~Kuhn, and JET
  EFDA~Contributors.
\newblock Kinetic simulations of the parallel transport in the jet scrape-off
  layer.
\newblock {\em Journal of Nuclear Materials}, 390-391, 06 2009.

\bibitem{Manfredi11}
Giovanni Manfredi, Sever~Adrian Hirstoaga, St{\'e}phane Devaux, Eva Havlickova,
  and David Tskhakaya.
\newblock Parallel transport in a tokamak scrape-off layer.
\newblock {\em 38th EPS Conference on Plasma Physics 2011, EPS 2011 -
  Europhysics Conference Abstracts}, 35, 06 2011.

\bibitem{Schneider92}
R~Schneider, D~Reiter, HP~Zehrfeld, B~Braams, Martine Baelmans, J~Geiger,
  H~Kastelewicz, J~Neuhauser, and R~Wunderlich.
\newblock B2-eirene simulation of asdex and asdex-upgrade scrape-off layer
  plasmas.
\newblock {\em Journal of nuclear materials}, 196:810--815, 1992.

\bibitem{Schneider06}
R~Schneider, X~Bonnin, K~Borrass, DP~Coster, H~Kastelewicz, D~Reiter,
  VA~Rozhansky, and BJ~Braams.
\newblock Plasma edge physics with b2-eirene.
\newblock {\em Contributions to Plasma Physics}, 46(1-2):3--191, 2006.

\bibitem{Rozhansky09}
V~Rozhansky, E~Kaveeva, P~Molchanov, I~Veselova, S~Voskoboynikov, D~Coster,
  G~Counsell, A~Kirk, S~Lisgo, MAST Team, et~al.
\newblock New b2solps5. 2 transport code for h-mode regimes in tokamaks.
\newblock {\em Nuclear Fusion}, 49(2):025007, 2009.

\bibitem{Stegmeir18}
Andreas Stegmeir, David Coster, Alexander Ross, Omar Maj, Karl Lackner, and
  Emanuele Poli.
\newblock Grillix: a 3d turbulence code based on the flux-coordinate
  independent approach.
\newblock {\em Plasma Physics and Controlled Fusion}, 60(3):035005, 2018.

\bibitem{Zholobenko19}
W~Zholobenko, A~Stegmeir, T~Body, A~Ross, P~Manz, O~Maj, D~Coster, F~Jenko,
  M~Francisquez, B~Zhu, et~al.
\newblock Thermal dynamics in the flux-coordinate independent turbulence code
  grillix.
\newblock {\em Contributions to Plasma Physics}, 60(5-6):e201900131, 2020.

\bibitem{Scott03}
Bruce~D Scott.
\newblock Computation of electromagnetic turbulence and anomalous transport
  mechanisms in tokamak plasmas.
\newblock {\em Plasma physics and controlled fusion}, 45(12A):A385, 2003.

\bibitem{Takizuka_2017}
T~Takizuka.
\newblock Kinetic effects in edge plasma: kinetic modeling for edge plasma and
  detached divertor.
\newblock {\em Plasma Physics and Controlled Fusion}, 59(3):034008, feb 2017.

\bibitem{Tronko16}
Natalia Tronko, Alberto Bottino, and Eric Sonnendr{\"u}cker.
\newblock Second order gyrokinetic theory for particle-in-cell codes.
\newblock {\em Physics of Plasmas}, 23(8):082505, 2016.

\bibitem{Sugama00}
H.~Sugama.
\newblock Gyrokinetic field theory.
\newblock {\em Physics of Plasmas}, 7:466, 2000.

\bibitem{Brizard20004816}
A.J. Brizard.
\newblock Variational principle for nonlinear gyrokinetic {Vlasov-Maxwell}
  equations.
\newblock {\em Physics of Plasmas}, 7(12):4816--4822, 2000.

\bibitem{Chang17}
Choong~Seock Chang, S~Ku, Alberto Loarte, Vassili Parail, Florian Koechl,
  Michele Romanelli, Rajesh Maingi, J-W Ahn, T~Gray, J~Hughes, et~al.
\newblock Gyrokinetic projection of the divertor heat-flux width from present
  tokamaks to iter.
\newblock {\em Nuclear Fusion}, 57(11):116023, 2017.

\bibitem{DominskiPoP2024_1}
J.~Dominski, C.S. Chang, R.~Hager, S.~Ku, E.S. Yoon, and V.~Parail.
\newblock {Neoclassical transport of tungsten ion bundles in total-f
  neoclassical gyrokinetic simulations of a whole-volume JET-like plasma}.
\newblock {\em Physics of Plasmas}, 31(3):032303, 03 2024.

\bibitem{DominskiPoP2024_2}
J.~Dominski, C.S. Chang, R.~Hager, P.~Helander, S.~Ku, and E.S. Yoon.
\newblock {Study of up–down poloidal density asymmetry of high-$Z$ impurities
  with the new impurity version of XGCa}.
\newblock {\em Journal of Plasma Physics}, 85(5):905850510, 2019.

\bibitem{Hager_2019}
R.~Hager, C.S. Chang, N.M. Ferraro, and R.~Nazikian.
\newblock Gyrokinetic study of collisional resonant magnetic perturbation
  (rmp)-driven plasma density and heat transport in tokamak edge plasma using a
  magnetohydrodynamic screened rmp field.
\newblock {\em Nuclear Fusion}, 59(12):126009, sep 2019.

\bibitem{Bott15}
A.~Bottino and E.~Sonnendr{\"u}cker.
\newblock {Monte Carlo particle-in-cell methods for the simulation of the
  Vlasov--Maxwell gyrokinetic equations}.
\newblock {\em Journal of Plasma Physics}, 81(5), 2015.

\bibitem{Lee83}
WW~Lee.
\newblock Gyrokinetic approach in particle simulation.
\newblock {\em The Physics of Fluids}, 26(2):556--562, 1983.

\bibitem{KU2016467}
S.~Ku, R.~Hager, C.S. Chang, J.M. Kwon, and S.E. Parker.
\newblock A new hybrid-lagrangian numerical scheme for gyrokinetic simulation
  of tokamak edge plasma.
\newblock {\em Journal of Computational Physics}, 315:467--475, 2016.

\bibitem{Jenko01}
Frank Jenko and W~Dorland.
\newblock Nonlinear electromagnetic gyrokinetic simulations of tokamak plasmas.
\newblock {\em Plasma Physics and Controlled Fusion}, 43(12A):A141, 2001.

\bibitem{Hakim2020}
A.H. Hakim, N.R. Mandell, T.N. Bernard, M.~Francisquez, G.W. Hammett, and E.L.
  Shi.
\newblock Continuum electromagnetic gyrokinetic simulations of turbulence in
  the tokamak scrape-off layer and laboratory devices.
\newblock {\em Physics of Plasmas}, 27(4), 2020.
\newblock Cited by: 24; All Open Access, Bronze Open Access.

\bibitem{Krommes12}
John~A Krommes.
\newblock The gyrokinetic description of microturbulence in magnetized plasmas.
\newblock {\em Annual Review of Fluid Mechanics}, 44:175--201, 2012.

\bibitem{Tskhakaya07}
David Tskhakaya, K~Matyash, R~Schneider, and F~Taccogna.
\newblock The particle-in-cell method.
\newblock {\em Contributions to Plasma Physics}, 47(8-9):563--594, 2007.

\bibitem{Hockney88}
R.W. Hockney and J.W. Eastwood.
\newblock {\em Computer simulation using particles}.
\newblock CRC Press, 1988.

\bibitem{Birdsall04}
C.K. Birdsall and A.B. Langdon.
\newblock {\em Plasma physics via computer simulation}.
\newblock CRC Press, 2004.

\bibitem{Garbet10}
X.~Garbet, Y.~Idomura, L.~Villard, and T.H. Watanabe.
\newblock Gyrokinetic simulations of turbulent transport.
\newblock {\em Nuclear Fusion}, 50(4):043002, 2010.

\bibitem{Krommes07}
John~A Krommes.
\newblock Nonequilibrium gyrokinetic fluctuation theory and sampling noise in
  gyrokinetic particle-in-cell simulations.
\newblock {\em Physics of Plasmas}, 14(9):090501, 2007.

\bibitem{Bottino07}
A~Bottino, AG~Peeters, R~Hatzky, S~Jolliet, BF~McMillan, TM~Tran, and
  L~Villard.
\newblock Nonlinear low noise particle-in-cell simulations of electron
  temperature gradient driven turbulence.
\newblock {\em Physics of plasmas}, 14(1):010701, 2007.

\bibitem{Hatzky07}
Roman Hatzky, Axel K{\"o}nies, and Alexey Mishchenko.
\newblock Electromagnetic gyrokinetic pic simulation with an adjustable control
  variates method.
\newblock {\em Journal of Computational Physics}, 225(1):568--590, 2007.

\bibitem{Mishchenko04}
Alexey Mishchenko, Roman Hatzky, and Axel K{\"o}nies.
\newblock Conventional $\delta$f-particle simulations of electromagnetic
  perturbations with finite elements.
\newblock {\em Physics of plasmas}, 11(12):5480--5486, 2004.

\bibitem{Kleiber16}
R~Kleiber, R~Hatzky, A~K{\"o}nies, A~Mishchenko, and E~Sonnendr{\"u}cker.
\newblock An explicit large time step particle-in-cell scheme for nonlinear
  gyrokinetic simulations in the electromagnetic regime.
\newblock {\em Physics of Plasmas}, 23(3):032501, 2016.

\bibitem{Mishchenko17}
Alexey Mishchenko, Alberto Bottino, Roman Hatzky, Eric Sonnendr{\"u}cker, Ralf
  Kleiber, and Axel K{\"o}nies.
\newblock Mitigation of the cancellation problem in the gyrokinetic
  particle-in-cell simulations of global electromagnetic modes.
\newblock {\em Physics of Plasmas}, 24(8):081206, 2017.

\bibitem{Idomura14}
Yasuhiro Idomura.
\newblock Full-f gyrokinetic simulation over a confinement time.
\newblock {\em Physics of Plasmas}, 21(2):022517, 2014.

\bibitem{Xu07}
XQ~Xu, Z~Xiong, MR~Dorr, JA~Hittinger, K~Bodi, J~Candy, BI~Cohen, RH~Cohen,
  P~Colella, GD~Kerbel, et~al.
\newblock Edge gyrokinetic theory and continuum simulations.
\newblock {\em Nuclear fusion}, 47(8):809, 2007.

\bibitem{Scott10GEM}
Bruce~D Scott, A~Kendl, and T~Ribeiro.
\newblock Nonlinear dynamics in the tokamak edge.
\newblock {\em Contributions to Plasma Physics}, 50(3-5):228--241, 2010.

\bibitem{Lee18cogent}
Wonjae Lee, MA~Dorf, MR~Dorr, RH~Cohen, TD~Rognlien, JAF Hittinger, MV~Umansky,
  and SI~Krasheninnikov.
\newblock Verification of 5d continuum gyrokinetic code cogent: Studies of
  kinetic drift wave instability.
\newblock {\em Contributions to Plasma Physics}, 58(6-8):445--450, 2018.

\bibitem{Hakim14}
AH~Hakim, GW~Hammett, and EL~Shi.
\newblock On discontinuous galerkin discretizations of second-order
  derivatives.
\newblock {\em arXiv preprint arXiv:1405.5907}, 2014.

\bibitem{Shi17}
EL~Shi, GW~Hammett, T~Stoltzfus-Dueck, and A~Hakim.
\newblock Gyrokinetic continuum simulation of turbulence in a straight
  open-field-line plasma.
\newblock {\em Journal of Plasma Physics}, 83(3):905830304, 2017.

\bibitem{Shi18}
Eric~L Shi, Gregory~W Hammett, Timothy Stoltzfus-Dueck, and Ammar Hakim.
\newblock Full-f gyrokinetic simulation of turbulence in a helical
  open-field-line plasma.
\newblock {\em Physics of Plasmas}, 26(1):012307, 2018.

\bibitem{Michels20}
Dominik Michels, Andreas Stegmeir, Philipp Ulbl, Denis Jarema, and Frank Jenko.
\newblock Gene-x: A full-f gyrokinetic turbulence code based on the
  flux-coordinate independent approach.
\newblock {\em Computer Physics Communications}, 264:107986, 2021.

\bibitem{Boesl2019}
M.~Boesl, A.~Bergmann, A.~Bottino, D.~Coster, E.~Lanti, N.~Ohana, and F.~Jenko.
\newblock Gyrokinetic full-f particle-in-cell simulations on open field lines
  with {PICLS}.
\newblock {\em Physics of Plasmas}, 26(12):122302, 2019.

\bibitem{Boesl2020}
M.~Boesl, A.~Bergmann, A.~Bottino, S.~Brunner, D.~Coster, and F.~Jenko.
\newblock Collisional gyrokinetic full-f particle-in-cell simulations on open
  field lines with {PICLS}.
\newblock {\em Contributions to Plasma Physics}, page e201900117, 12 2019.

\bibitem{Loarte07}
A~Loarte, B~Lipschultz, AS~Kukushkin, GF~Matthews, PC~Stangeby, N~Asakura,
  GF~Counsell, G~Federici, A~Kallenbach, K~Krieger, et~al.
\newblock Power and particle control.
\newblock {\em Nuclear Fusion}, 47(6):S203, 2007.

\bibitem{PARKER1993}
S.E. Parker, R.J. Procassini, Birdsall C.K., and Cohen B.I.
\newblock A suitable boundary condition for bounded plasma simulation without
  sheath resolution.
\newblock {\em Journal of Computational Physics}, 104(1):41--49, 1993.

\bibitem{Tronko18}
N.~Tronko and C.~Chandre.
\newblock Second-order nonlinear gyrokinetic theory: from the particle to the
  gyrocentre.
\newblock {\em Journal of Plasma Physics}, 84(3), 2018.

\bibitem{Dubin83}
D.H.E. Dubin, J.A. Krommes, C.~Oberman, and W.W. Lee.
\newblock Nonlinear gyrokinetic equations.
\newblock {\em Physics of Fluids}, 26(12):3524--3535, 1983.

\bibitem{lantiCPC2020}
E.~Lanti, N.~Ohana, N.~Tronko, T.~Hayward-Schneider, A.~Bottino, B.~F.
  McMillan, A.~Mishchenko, A.~Scheinberg, A.~Biancalani, P.~Angelino,
  S.~Brunner, J.~Dominski, P.~Donnel, C.~Gheller, R.~Hatzky, A.~Jocksch,
  S.~Jolliet, Z.~X. Lu, J.~P.~Martin Collar, I~Novikau, E.~Sonnendruecker,
  T.~Vernay, and L.~Villard.
\newblock {Orb5: A global electromagnetic gyrokinetic code using the PIC
  approach in toroidal geometry}.
\newblock {\em {COMPUTER PHYSICS COMMUNICATIONS}}, {251}, {JUN} {2020}.

\bibitem{HatzkyJPP2019}
R.~Hatzky, R.~Kleiber, A.~Könies, A.~Mishchenko, M.~Borchardt, A.~Bottino, and
  E.~Sonnendrücker.
\newblock Reduction of the statistical error in electromagnetic gyrokinetic
  particle-in-cell simulations.
\newblock {\em Journal of Plasma Physics}, 85(1):905850112, 2019.

\bibitem{DominskiPoP2018}
J.~Dominski, S.~Ku, and C.S. Chang.
\newblock {Gyroaveraging operations using adaptive matrix operators}.
\newblock {\em Physics of Plasmas}, 25(5):052304, 05 2018.

\bibitem{Bottino2022}
A.~Bottino, M.V. Falessi, T.~Hayward-Schneider, A.~Biancalani, S.~Briguglio,
  R.~Hatzky, Ph. Lauber, A.~Mishchenko, E.~Poli, B.~Rettino, F.~Vannini,
  X.~Wang, and F.~Zonca.
\newblock Time evolution and finite element representation of phase space zonal
  structures in orb5.
\newblock {\em Journal of Physics: Conference Series}, 2397(1):012019, dec
  2022.

\bibitem{Stier2024}
A.~Stier, A.~Bottino, M.~Boesl, M.~Campos~Pinto, A.~Bergmann, M.~Murugappan,
  S.~Brunner, L.~Villard, and F.~Jenko.
\newblock Verification of the fourtier-enhanced 3d finite element poisson
  solver of the gyrokinetic full-f code picls.
\newblock {\em Computer Physics Communications}, 229:109155, 2024.

\bibitem{laug}
E.~Anderson, Z.~Bai, C.~Bischof, S.~Blackford, J.~Demmel, J.~Dongarra,
  J.~Du~Croz, A.~Greenbaum, S.~Hammarling, A.~McKenney, and D.~Sorensen.
\newblock {\em {LAPACK} Users' Guide}.
\newblock Society for Industrial and Applied Mathematics, Philadelphia, PA,
  third edition, 1999.

\bibitem{Oberkampf10}
William~L Oberkampf and Christopher~J Roy.
\newblock {\em Verification and validation in scientific computing}.
\newblock Cambridge University Press, 2010.

\bibitem{parker93sheath}
S.E. Parker, R.J. Procassini, C.K. Birdsall, and B.I. Cohen.
\newblock A suitable boundary condition for bounded plasma simulation without
  sheath resolution.
\newblock {\em Journal of Computational Physics}, 104(1):41--49, 1993.

\bibitem{TOGO2016109}
Satoshi Togo, Tomonori Takizuka, Makoto Nakamura, Kazuo Hoshino, Kenzo Ibano,
  Tee~Long Lang, and Yuichi Ogawa.
\newblock Self-consistent treatment of the sheath boundary conditions by
  introducing anisotropic ion temperatures and virtual divertor model.
\newblock {\em Journal of Computational Physics}, 310:109--126, 2016.

\bibitem{HATZKY2006325}
R.~Hatzky.
\newblock Domain cloning for a particle-in-cell (pic) code on a cluster of
  symmetric-multiprocessor (smp) computers.
\newblock {\em Parallel Computing}, 32(4):325--330, 2006.

\bibitem{Biancalani2016}
A.~Biancalani, A.~Bottino, S.~Briguglio, A.~Könies, Ph. Lauber, A.~Mishchenko,
  E.~Poli, B.~D. Scott, and F.~Zonca.
\newblock {Linear gyrokinetic particle-in-cell simulations of Alfvén
  instabilities in tokamaks}.
\newblock {\em Physics of Plasmas}, 23(1):012108, 01 2016.

\bibitem{Lysack23}
R.L. Lysak.
\newblock Kinetic alfven waves and auroral particle acceleration: a review.
\newblock {\em Rev. Mod. Plasma Phys.}, page~6, 7 2023.

\end{thebibliography}
\bibliographystyle{unsrt} 

\end{document}